%% file: main.tex
\documentclass[acmtog]{acmart}

\usepackage{url}

\newcommand{\lz}[1]{{\color{black} #1}}

\newcommand{\zj}[1]{{\color{black} #1}}

\copyrightyear{2025}
\acmYear{2025}
\setcopyright{acmlicensed}

\begin{document}

\title{From Rigging to Waving: 3D-Guided Diffusion for Natural Animation of Hand-Drawn Characters}

\author{Jie Zhou$^\ast$}
\orcid{0000-0002-3836-4163}
\email{jzhou67-c@my.cityu.edu.hk}
\affiliation{
    \institution{City University of Hong Kong}
    \country{China}
}

\author{Linzi Qu$^\ast$}
\orcid{0000-0002-8731-5501}
\email{linziqu2-c@my.cityu.edu.hk}
\affiliation{
    \institution{City University of Hong Kong}
    \country{China}
}

\author{Miu-Ling Lam$^\dagger$}
\orcid{0000-0002-5333-7454}
\email{miu.lam@cityu.edu.hk}
\affiliation{
    \institution{City University of Hong Kong}
    \country{China}
}

\author{Hongbo Fu$^\dagger$}
\orcid{0000-0002-0284-726X}
\email{fuplus@gmail.com}
\affiliation{
    \institution{Hong Kong University of Science and Technology}
    \country{China}
}
\thanks{$^\ast$ Equal Contributions.}
\thanks{$^\dagger$ Corresponding authors.}
\renewcommand\shortauthors{Zhou and Qu et al.}

\begin{abstract}
\input{files/abstract}

\end{abstract}

%
%

\ccsdesc[500]{Computing methodologies~Animation}
\ccsdesc[300]{Computing methodologies~Non-photorealistic rendering}
%
%

\keywords{Character Animation, Video Diffusion Model, Secondary Motion, Skeletal Animation}

\input{figures/fig_teaser}

\maketitle

\input{files/introduction}

\input{files/related_work}

\input{files/method}

\input{files/evaluation}

\input{files/application}

\input{files/conclusion}


\bibliographystyle{ACM-Reference-Format}
\bibliography{files/reference}

\end{document}


\title{Supplemental Materials for "From Rigging to Waving: 3D-Guided Diffusion for Natural Animation of Hand-Drawn Characters"}

\author{Jie Zhou$^\ast$}
\orcid{0000-0002-3836-4163}
\email{jzhou67-c@my.cityu.edu.hk}
\affiliation{
    \institution{City University of Hong Kong}
    \country{China}
}

\author{Linzi Qu$^\ast$}
\orcid{0000-0002-8731-5501}
\email{linziqu2-c@my.cityu.edu.hk}
\affiliation{
    \institution{City University of Hong Kong}
    \country{China}
}

\author{Miu-Ling Lam$^\dagger$}
\orcid{0000-0002-5333-7454}
\email{miu.lam@cityu.edu.hk}
\affiliation{
    \institution{City University of Hong Kong}
    \country{China}
}

\author{Hongbo Fu$^\dagger$}
\orcid{0000-0002-0284-726X}
\email{fuplus@gmail.com}
\affiliation{
    \institution{Hong Kong University of Science and Technology}
    \country{China}
}
\thanks{$^\ast$ Equal Contributions.}
\thanks{$^\dagger$ Corresponding Authors.}

\maketitle

\zj{
\section{Notation Table}
Table \ref{tab:notation} lists the symbols used in this paper along with their corresponding descriptions for the reader's reference.
\begin{table}[h]
    \centering
    \renewcommand\arraystretch{1.25}
    \caption{Notation Table}
    \label{tab:notation}
    \begin{tabular}{@{}llp{5cm}@{}}
        \toprule
        Notation & Description \\ \midrule
        $I_{\text{ref}}$ & Reference image \\
        $I^\text{nc}_{\text{ref}}$ & Contour-free reference image \\
        $\mathcal{P}$ & Target 3D motion \\
        $\{\hat{I}_{1:N}\}$ & Generated 3D character animation \\
        $S_{\text{front}}$ & Hair-body segmentation map (front-view) \\
        $S_{\text{right}}$ & Hair-body segmentation map (right-view) \\ 
        $M_{\text{front}}^{\text{SDI}}$ & SDI mask (front-view) \\
        $M_{\text{back}}^{\text{SDI}}$ & SDI mask (back-view) \\
        $\mathcal{G}$ & 3D textured geometry reconstructed from $I_{\text{ref}}$ \\
        $E(\cdot)$ & VAE encoder\\
        $D(\cdot)$ & VAE decoder\\
        $\{P_{1:N}\}$ & Pose guidance sequence\\
        $\{M_{1:N}^{\text{SDI}}\}$ & SDI mask guidance sequence\\
        $\{C_{1:N}\}$ & Coarse color guidance sequence \\
        $\{C_{1:N}^\text{masked}\}$ & Masked coarse color guidance sequence \\
        $P_{\text{ref}}$ & Reference pose extracted from $I_{\text{ref}}$\\
        $\mathscr{E}_{\text{coarse}}$ & Coarse prior encoder \\
        $u_\theta$ & Native pre-trained UniAnimate\\
        $v_\theta$ & Our domain-adapted model\\
        $\{\hat{z}_{1:N,0}^{t, u_\theta}\}$ & Noise-free latent estimates from $u_\theta$\\
        $\{\hat{z}_{1:N,0}^{t, v_\theta}\}$ & Noise-free latent estimates from $v_\theta$\\
        $\{\hat{z}_{1:N,0}^{t, \text{blend}}\}$ &Blended noise-free latent estimates\\
        T & The total number of denoising steps\\
        $\tau_2$ & The starting step of SDI\\
        $\tau_1$ & The ending step of SDI\\
        $\alpha$ & The percentage to control $\tau_2$ where $\tau_2=\alpha \cdot T$\\
        $\beta$ & The percentage to control $\tau_1$ where $\tau_1=\beta \cdot T$\\

        \bottomrule
    \end{tabular}
\end{table}
}

\section{Pose Guidance Extraction}
We describe how we extract the pose sequence $\{P_{1:N}\}$ and the reference pose $P_{\text{ref}}$ from the animated character. We adopt the OpenPose \cite{cao2019openpose} $18$-keypoint format, excluding the hand and feet joints due to the abstract nature of hand-drawn characters.
Since the animated 3D skeleton generated from Mixamo does not include facial keypoints, we employ an approximate method to locate them: we predict the 2D positions of $5$ facial keypoints (nose, eyes, ears) from $I_{\text{ref}}$ using X-Pose \cite{yang2024x} and then back-project these points onto the surface of $\mathcal{G}$ to obtain their depth values. By this, we get the 3D positions of facial keypoints in the character's rest pose. Inspired by Astropulse's tool \cite{mixamotoopenpose}, we assume that the character's head is a rigid body, and calculate the new positions of the facial keypoints based on the orientation of the head in each frame and relative positions between the head and facial keypoints. For the other $13$ body keypoints, we directly extract their per-frame 3D position from the animated 3D skeleton. Finally, we sort the $17$ limbs formed by these $18$ keypoints according to depth to ensure that their 2D projections maintain the correct occlusion order.

\zj{
\section{Algorithm for SDI}
The algorithm \ref{alg:SDI} describes the process for generating the synthesized video using reference images and guidance sequences.

\begin{algorithm}
    \caption{Generation Process with Secondary Dynamics Injection (SDI)}
    \label{alg:SDI}
    \small
    \begin{algorithmic}[]
        \STATE \textbf{Input:} 
        \begin{itemize}
            \item Total timesteps $T$ 
            \item Thresholds $\tau_2$, $\tau_1$
            \item Reference image $I_\text{ref}$
            \item Contour-free reference image $I^\text{nc}_\text{ref}$ 
            \item Reference pose $P_{\text{ref}}$ 
            \item Pose guidance sequence $\{P_{1:N}\}$ 
            \item Masked coarse color guidance sequence $\{C^{\text{masked}}_{1:N}\}$ 
            \item SDI mask guidance sequence $\{M_{1:N}^{\text{SDI}}\}$
        \end{itemize}
        \STATE \textbf{Initialization:} Input random noise sequence $\{z_{1:N,T}\}$
        \STATE \textbf{Output:} Synthesized video $\{\hat{I}_{1:N}\}$

        
        \STATE \textbf{Stage 1:}
        \FOR{each timestep $t \in [T, \tau_2)$}
            \STATE $ \{z_{1:N, t-1}\} \gets v_{\theta}(I^\text{nc}_\text{ref}, \{z_{1:N, t}\}, P_{\text{ref}}, \{P_{1:N}\}, \{C^{\text{masked}}_{1:N}\}, \{M_{1:N}^{\text{SDI}}\})$
        \ENDFOR

        
        \STATE \textbf{Stage 2:}
        \STATE \textbf{Initialization:} $\{z_{1:N, \tau_2}^{v_\theta}\}=\{z_{1:N, \tau_2}\}, \{z_{1:N, \tau_2}^{u_\theta}\}=\{z_{1:N, \tau_2}\}$
        \FOR{each timestep $t \in [\tau_2, \tau_1)$} 
            \STATE $\{z_{1:N, t-1}^{v_\theta}\} \gets v_{\theta}(I^\text{nc}_\text{ref}, \{z_{1:N, t}^{v_\theta}\}, P_{\text{ref}}, \{P_{1:N}\}, \{C^{\text{masked}}_{1:N}\}, \{M_{1:N}^{\text{SDI}}\})$
            \STATE $\{z_{1:N, t-1}^{u_\theta}\} \gets u_{\theta}(I^\text{nc}_\text{ref}, \{z_{1:N, t}^{u_\theta}\}, P_{\text{ref}}, \{P_{1:N}\})$

            \STATE \textbf{Extract the noise-free latent estimates:}
            \[
            \{\hat{z}_{1:N,0}^{t-1, v_\theta}\} \gets \{z_{1:N, t-1}^{v_\theta}\}, \quad 
            \{\hat{z}_{1:N,0}^{t-1, u_\theta}\} \gets \{z_{1:N, t-1}^{u_\theta}\}
            \]

            \STATE \textbf{Perform blending using the SDI mask:}
            \[
                \begin{aligned}
                    \{\hat{z}_{1:N, 0}^{t-1, \text{blend}}\} &= (1-\{M^\text{SDI}_{1:N, \text{down}}\}) \cdot \{\hat{z}_{1:N,0}^{t-1, v_\theta}\} \\
                    &\quad + \{M^\text{SDI}_{1:N, \text{down}}\} \cdot \{\hat{z}_{1:N,0}^{t-1, u_\theta}\}
                    \end{aligned}
            \]

            \STATE \textbf{Guide the denoising direction:}
            \[
            \{z_{1:N, t-1}^{v_\theta}\} \gets \{\hat{z}_{1:N, 0}^{t-1, \text{blend}}\}, 
            \{z_{1:N, t-1}^{u_\theta}\} \gets \{\hat{z}_{1:N, 0}^{t-1, \text{blend}}\}
            \]
            
        \ENDFOR

        \STATE \textbf{Inpaint coarse frames using Poisson blending:}
        \[
            \{C^\text{inpainted}_{1:N}\} \gets \text{PoissonBlend}(\{C^\text{masked}_{1:N}\}, D(\hat{z}_{1:N, 0}^{\tau_1, \text{blend}})) 
        \]
        \[
            \{M_{1:N}^{\text{black}}\} \gets \text{Placeholder: Full Black Mask }
        \]

        
        \STATE \textbf{Stage 3:}
        \FOR{each timestep $t \in [T, 1]$} 
            \STATE $\{z_{1:N, t-1}\} \gets v_{\theta}(I_\text{ref}, \{z_{1:N, t}\}, P_{\text{ref}}, \{P_{1:N}\}, \{C^{\text{inpainted}}_{1:N}\}, \{M_{1:N}^{\text{black}}\})$
        \ENDFOR


        \STATE \textbf{Output:} $\{\hat{I}_{1:N}\} \gets  D(\{z_{1:N, 0}\})$
    \end{algorithmic}
\end{algorithm}
}

\section{Perceptual User Study}
We performed a user study to assess the perceptual quality of our proposed method (Ours) against the baseline approaches. The evaluation was based on three key metrics:
(1) Pose Consistency (PC): Measures the alignment between target poses and generated results. (2) Style Preservation (SP): Evaluates how well the original reference style is retained, especially for the contour regions. (3) Motion Naturalness (MN): Assesses the realism and plausibility of motion dynamics, with emphasis on secondary motion. 
The study was conducted through an online questionnaire, featuring 20 randomly ordered result sets spanning a diverse range of character styles (e.g., cartoon characters and anthropomorphic animals) and motion styles (e.g., casual walking, athletic jumps, and expressive dances) to ensure broad stylistic coverage. A total of 50 participants were recruited for the study. They were tasked with identifying and selecting all methods that aligned closely with each specific evaluation standard (PC, SP, MN).
Table \ref{tab:user} presents the voting results, indicating that our method was consistently the most frequently selected among the three evaluation criteria.
This preference, expressed by human subjects, highlights the effectiveness of our approach in animating stylized characters with natural motions and ensuring character-specific details.
Since both our method and DrawingSpinUp are based on skeletal animation, the pose consistency metrics are comparable. However, when combined with diffusion priors, our method significantly surpasses DrawingSpinUp in terms of stylistic consistency and motion naturalness.
\vspace{-3mm}
\begin{table}[htbp]
\begin{center}
\scalebox{0.9}{
\begin{tabular}{cccc}
    \hline
	\multicolumn{1}{c}{\textbf{Method}}
        & \multicolumn{1}{c}{\textbf{PC $\uparrow$}}
        & \multicolumn{1}{c}{\textbf{SP $\uparrow$}}
	& \multicolumn{1}{c}{\textbf{MN $\uparrow$}} \\
	\hline
    
    AnimateAnyone & 6.0 & 8.0 & 5.0  \\ 
    MikuDance & 10.8 & 6.6 & 7.7  \\
    UniAnimate & 10.8 & 9.9 & 8.8  \\
    UniAnimate* & 21.5 & 20.7 & 20.1  \\
    DrawingSpinUp  & 25.4 & 26.7 & 24.4  \\
    \hline
    Ours  & \textbf{25.5} & \textbf{28.1} & \textbf{34.0}  \\
    \hline
    
\end{tabular}
}
\end{center}
\caption{Summary of the voting results from the user study.}
\vspace{-9mm}
\label{tab:user}
\end{table}

\section{SDI Masks}
\input{figures/fig_SDI_mask_show}
Fig. \ref{fig:SDI_mask} shows the SDI masks corresponding to all the examples used in the paper.
The ranges of the SDI masks are relatively flexible. Users can choose to mask only the hair ends or the entire hair, leading to different final redraws. 
The larger the range of the SDI mask, the greater the enhancement range of the diffusion model; however, this may also introduce increased geometric distortion. This creates a trade-off that needs to be balanced.

\bibliographystyle{ACM-Reference-Format}
\bibliography{files/reference}

%% file: files/abstract.tex
Hand-drawn character animation is a vibrant research area in computer graphics and presents unique challenges in achieving geometric consistency while conveying expressive motion details.
Traditional skeletal animation methods maintain geometric consistency but often struggle with complex non-rigid elements like flowing hair and skirts, resulting in unnatural deformation and missing secondary dynamics.
In contrast, video diffusion models effectively synthesize physically plausible dynamics, but exhibit real-human-like characteristics and geometric distortions when applied to stylized drawings due to the domain gap.
In this work, we propose a novel hybrid animation system that integrates the strengths of skeletal animation and video diffusion priors. 
The core idea is to first generate coarse images from characters retargeted with skeletal animations for geometric consistency guidance, and then enhance these images in terms of texture details and secondary dynamics using video diffusion priors.
We formulate the enhancement of coarse images as an inpainting task and propose a domain-adapted diffusion model to refine user-masked regions requiring improvement, particularly those involving secondary dynamics.
To further enhance motion realism, we propose a Secondary Dynamics Injection (SDI) strategy during the denoising process to incorporate latent features from a pre-trained diffusion model enriched with human motion priors.
Additionally, to address unnatural deformation artifacts caused by the integrated hair-body geometry in low-poly single-mesh character modeling, we introduce a Hair Layering Modeling (HLM) technique that employs segmentation maps to separate hair from the body in implicit fields, enabling more natural animation of challenging long-hair characters.
Through extensive experiments, we demonstrate that our system outperforms state-of-the-art works in both quantitative and qualitative evaluations.
Please refer to this anonymous GitHub repository (\url{https://anonymous.4open.science/r/From-Rigging-to-Waving-405C}) for our code, data and results.

%% file: figures/fig_teaser.tex
\begin{teaserfigure}
  \centering
  \includegraphics[width=1\textwidth]{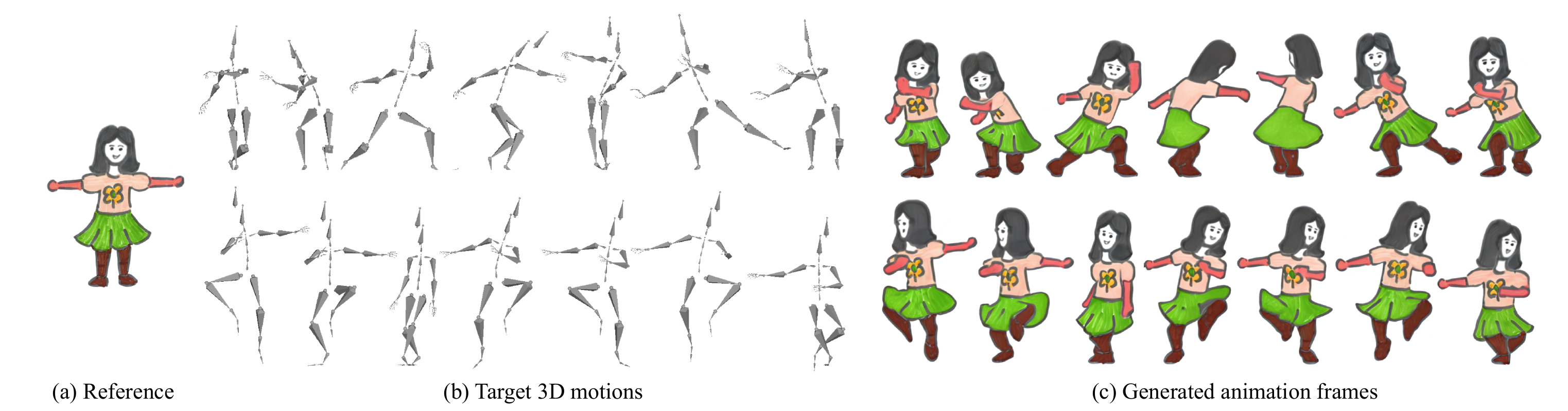}
    \vspace{-6mm}
  \caption{Our system produces natural character animations (Right) given single input drawings (Left) and target 3D motions (Middle).}
  \label{fig:teaser}
\end{teaserfigure}

%% file: files/introduction.tex
\section{Introduction}
Hand-drawn character animation has emerged as a captivating research area in computer graphics. Breathing life into static hand-drawn characters opens doors to innovation in entertainment, education, and more, creating interactive and immersive experiences. Common techniques for this task can be categorized into two main approaches: traditional skeletal animation and pose-controllable video diffusion models. However, both face significant challenges in simultaneously maintaining geometric consistency and conveying expressive motion details.

Traditional skeletal animation methods \cite{hornung2007character, smith2023method} embed hand-drawn characters as 2D meshes, rig them, and employ as-rigid-as-possible (ARAP) shape manipulation \cite{igarashi2005rigid} to repose character meshes, ensuring identity consistency, but they are limited to 2D planar motions. DrawingSpinUp \cite{zhou2024drawingspinup} improves this process by reconstructing 3D models from character drawings prior to rigging, supporting 3D motions while maintaining geometric consistency.
However, these methods struggle to animate characters with complex non-rigid elements, such as long hair or loose clothing. 
Due to the limitations of current single-to-3D generation methods \cite{liu2023one, tang2024dreamgaussian, long2024wonder3d, wang2024crm}, the 3D characters reconstructed from a single image often exhibit low-poly single-mesh geometry. Consequently, shoulder-length and longer hair frequently adheres to the neck and shoulders, leading to unnatural deformation (see Fig. \ref{fig:existing_problem} (a)-(b)).
Additionally, skeletal animation focuses on primary motions and thus fails to convey realistic secondary dynamics (such as subtle movements of hair and clothing). While artists can achieve realistic visual details through more complex rigs, hierarchical modeling, or meticulous adjustments of material parameters, these manual processes are labor-intensive and demand specialized expertise.

\input{figures/fig_existing_problem}

Recently, pose-controllable video diffusion models \cite{hu2024animate, zhu2024champ, wang2024unianimate, tan2024animate} have shown remarkable capabilities in generating dynamic motions for various identities within the domain of real human data. These models inherently learn physics-aware motion priors, effectively capturing both primary and secondary motions from extensive real human motion videos. Unfortunately, directly applying these models to hand-drawn character drawings often results in outputs that overlook specific artistic details, exhibit distorted appearances, particularly in regions close to the contour of the character drawing (see Fig. \ref{fig:existing_problem} (c)), and synthesize real-human-like characteristics (see Fig. \ref{fig:existing_problem} (d)).
These issues stem from the domain gap (e.g., shape proportion, contour lines, exaggerated motions) between hand-drawn characters and real humans. Although training a video diffusion model specifically for hand-drawn character animation could be a potential solution, collecting a substantial dataset of hand-drawn character animations with realistic secondary motion is impractical.
Building on these observations, we recognize a potential complementarity between skeletal animation and video diffusion models. We thus propose a novel hybrid animation system that integrates the strengths of both approaches, enabling the generation of diverse stylized hand-drawn character animations with natural motions.

Specifically, we employ skeletal animation to generate coarse images from retargeted characters, ensuring geometric consistency for stable animation generation. We formulate the enhancement of coarse images as an inpainting task and introduce a domain-adapted diffusion model to refine them primarily from two aspects: appearance details and secondary dynamics. To bridge the gap between real humans and hand-drawn characters, we construct a small-scale animation dataset focusing on primary motions to enhance texture details and contour lines. However, since this dataset emphasizes primary motions, the domain-adapted diffusion model might neglect secondary motions during animation generation. To address this limitation, we propose a Secondary Dynamics Injection (SDI) strategy during the denoising process, which enhances the naturalness of motion by leveraging a pre-trained diffusion model to guide the denoising direction.
This guidance is achieved by blending the latents from our domain-adapted diffusion model and the pre-trained one, using user-provided SDI masks.
Additionally, to address unnatural deformation artifacts caused by the integrated hair-body geometry in low-poly single-mesh character modeling, we introduce a Hair Layering Modeling (HLM) method that uses segmentation maps to separate hair from the body in implicit fields, allowing our system to animate challenging long-hair characters naturally.
The results of comprehensive experiments and a perceptual user study demonstrate that our system generates high-fidelity stylized animations that maintain geometric consistency while enhancing dynamic details. This underscores its superior performance compared to state-of-the-art methods. Furthermore, we highlight the potential of our system to facilitate a user-friendly application for editing existing animations.

%% file: figures/fig_existing_problem.tex
\begin{figure}[t]
    \centering
    \includegraphics[width=0.5\textwidth]{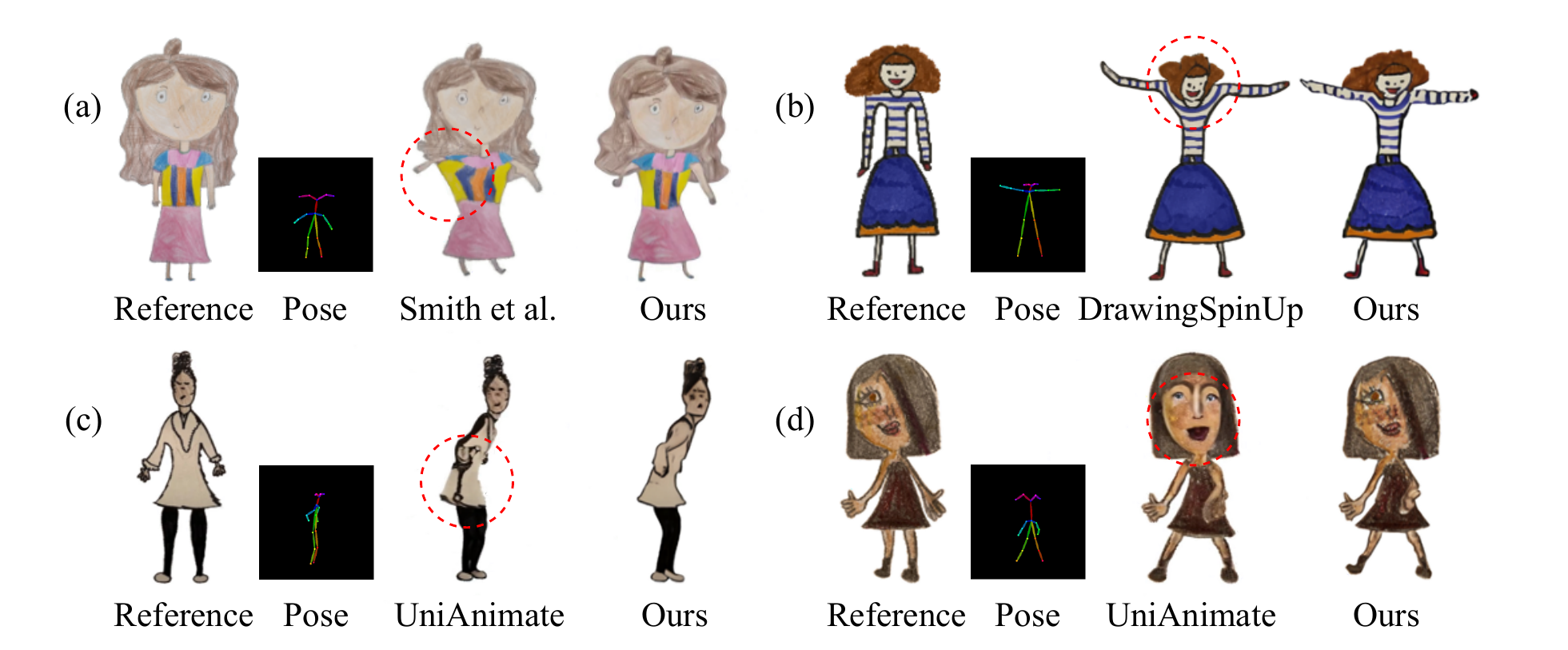}
    \vspace{-6mm}
    \caption{Illustration of issues with existing methods: (a)-(b) unnatural deformations; (c) incorrect contours; (d) inconsistent identity.}
    \vspace{-6mm}
    \label{fig:existing_problem}
\end{figure}

%% file: files/related_work.tex
\input{figures/fig_pipeline}

\section{Related Work}

\subsection{Hand-Drawn Character Animation}
The task of character drawing animation has been studied for a long time. 2D animation methods \cite{hornung2007character, smith2023method} typically project 3D motions onto the image plane to animate a 2D character using As-Rigid-As-Possible (ARAP) deformation \cite{igarashi2005rigid}. However, these methods are limited to generating results from a preset viewpoint. 3D animation methods \cite{weng2019photo, zhou2024drawingspinup} typically reconstruct 3D geometries as proxies from drawings and perform skeletal animation. However, since these methods deform the entire character as a single 2D/3D mesh, they demonstrate only a single primary dynamic effect and struggle to animate complex characters with long hair or loose clothing.

Some multi-layered methods \cite{willett2017secondary, fan2018tooncap}add 2D secondary motions on hair and clothing by deforming a 2D layered puppet. To achieve 3D secondary dynamics, Jain et al. \shortcite{jain2012three} propose using 3D proxies to simulate physically driven clothes for a 2D hand-drawn character. This approach relies on a 3D simulation engine, such as Maya, and is notably time-consuming. Zhang et al. \shortcite{zhang2020complementary} propose the concept of complementary dynamics to enhance rigged animations with detailed elastodynamics, though this incurs a computational cost. Benchekroun et al. \shortcite{benchekroun2023fast} further introduce a reduced-space elastodynamic solver to improve performance. However, these methods often result in a rubbery elasticity effect, suitable for flesh and skin but not for hair and fabrics. 
PhysAnimator \cite{xie2025physanimator} combines physics-based simulations with data-driven generative models to create intermediate frames between keyframes. While it introduces expressive, data-driven dynamics, it is limited to 2D representations and cannot fully capture 3D effects. In contrast, our method integrates 3D skeletal animation into a domain-adapted video diffusion model to produce high-fidelity stylized animations with 3D geometric consistency and secondary dynamic details.

\subsection{Pose-Controlled Video Diffusion Models}
Recently, with the advancement of large generative models \cite{blattmann2023stable, wang2024recipe}, pose-controlled video diffusion models have achieved remarkable results in generating natural physical dynamics. DisCo \cite{wang2024disco} pioneers this approach by incorporating a Pose ControlNet and a CLIP image encoder with a denoising UNet, separately guiding pose generation and encoding human semantics.
To further preserve identity consistency, methods such as \cite{hu2024animate, xu2024magicanimate, zhu2024champ} extract fine-grained identity features using a reference UNet that is architecturally aligned with the denoising UNet. These features are then injected into the denoising UNet through attention mechanisms.
UniAnimate \cite{wang2024unianimate} simplifies the model architecture by mapping the reference image, pose guidance, and noise into a unified feature space within a single video diffusion framework, thereby effectively reducing model complexity.
However, due to the domain gap between hand-drawn characters and real humans, most of the aforementioned approaches, trained on real-world human video datasets \cite{zablotskaia2019dwnet, jafarian2021learning}, fail to generalize effectively to diverse stylized characters.
It is impractical to prepare a large-scale dataset for these stylized character animations to address the domain gap.
Animate-X \cite{tan2024animate} attributes this limitation to an insufficient understanding of driving video motion patterns and misalignment with reference appearances. Thus, Animate-X introduces a Pose Indicator to extract motion patterns through both implicit and explicit mechanisms. Nevertheless, it overlooks the semantic understanding of reference images, and training on human videos struggles to establish correct contour relationships for characters in drawing styles.
To bridge this gap, MikuDance \cite{zhang2024mikudance} proposes a mixed-control diffusion module that implicitly aligns the scale and body shape of stylized characters with motion guidance. However, its dataset is heavily biased toward a specific anime style, which limits its generalization to a wider range of stylized characters.

Unlike the above approaches, we construct an explicit 3D model for an input hand-drawn character and use it to render guidance for directing the diffusion model's generation process. The generation task is repurposed as an inpainting task, where the rendered guidance ensures consistency in identity and pose, even under complex 3D motions. Additionally, our approach reduces the reliance on large-scale datasets for model training.

%% file: figures/fig_pipeline.tex
\begin{figure*}[h]
    \centering
    \includegraphics[width=1\textwidth]{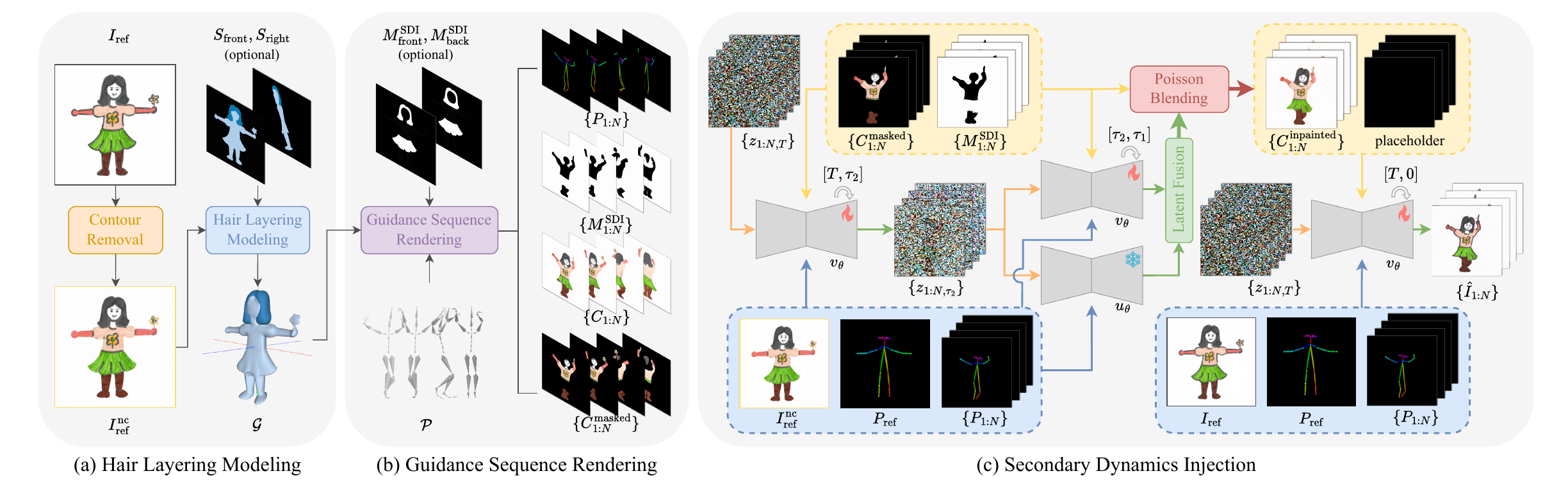}
    \vspace{-6mm}
    \caption{{An illustration of our pipeline, which consists of three main parts. 
    (a) Given a hand-drawn character image $I_{ref}$, we first remove the contour to obtain $I_{\text{ref}}'$ and then construct a hair-layered model $\mathcal{G}$.
    (b) According to the target 3D motion $\mathcal{P}$ and user-specified \zj{SDI} masks ($M_{\text{front}}^{\text{SDI}}$, $M_{\text{back}}^{\text{SDI}}$), we render several guidance sequences (pose $\{P_{1:N}\}$, SDI mask $\{M_{1:N}^{\text{SDI}}\}$, coarse color $\{C_{1:N}\}$, and masked coarse color $\{C^{\text{masked}}_{1:N}\}$) after rigging $\mathcal{G}$.
    (c) During the inference process, based on those guidance sequences, we begin denoising using $I^\text{nc}_\text{ref}$ via our domain-adapted model $v_{\theta}$ during the denoising steps $[T, \tau_2]$. 
    Next, still using $I^\text{nc}_\text{ref}$, we fuse the latents from $v_{\theta}$ with those from the pre-trained model $u_{\theta}$ during the denoising steps $[\tau_2, \tau_1]$ to optimize the denosing direction with natural motion. At the denoising step $\tau_1$, we estimate the output video and enhance $\{C^{\text{masked}}_{1:N}\}$ to create the inpainted coarse guidance $\{C^{\text{inpainted}}_{1:N}\}$ via Poisson Blending. Finally, guided by the $\{C^{\text{inpainted}}_{1:N}\}$ and $I_\text{ref}$, we further denoise from scratch to achieve the final results $\{\hat{I}_{1:N}\}$.}}
    \label{fig:pipeline}
\end{figure*}

%% file: files/method.tex
\section{Method}
Given an input hand-drawn character as a reference image $I_{\text{ref}}$ (we assume that the character in the reference image is approximately in a frontal A/T pose), \zj{manually-processed hair-body segmentation maps provided by the user (optional for long-hair cases)} and a target 3D motion $\mathcal{P}$, our system is designed to generate a vivid 3D character animation $\{\hat{I}_{1:N}\}$.  
An overview of our system is shown in Fig. \ref{fig:pipeline}. 
During the image-to-3D process, we design a Hair Layering Modeling (HLM) method based on hair-body segmentation maps ($S_{\text{front}}$, $S_{\text{right}}$) to deal with challenging characters with long hair (Section \ref{sec:hlm}). For characters without long hair, this step is optional.
Users can optionally specify areas requiring improvement,  particularly for dynamics enhancement, through \zj{SDI masks ($M_{\text{front}}^{\text{SDI}}$, $M_{\text{back}}^{\text{SDI}}$)}. 
Then our system reconstructs a 3D textured geometry $\mathcal{G}$ and renders three types of guidance sequences (Section \ref{sec:recon}), which provide essential 3D geometric consistency for subsequent animation generation.
We design a domain-adapted video diffusion model to translate rendered coarse color frames into stylized ones (Section \ref{sec:refine}). During inference, we employ a Secondary Dynamics Injection (SDI) strategy to further inject dynamic motions into the masked regions (Section \ref{sec:sec}).

\input{figures/fig_hlm}

\subsection{Preliminaries} \label{sec:pre}
Considering the computation cost, recent diffusion models \cite{rombach2022high, blattmann2023stable} operate within the compressed latent space of a pre-trained variational autoencoder \lz{(VAE)}, denoted as $E(\cdot)$ and $D(\cdot)$. Starting with the encoded latent representation $z_0=E(I)$ of an image $I$, the forward process $q(z_t|z_0, t)$ gradually corrupts the initial latent $z_0$ into Gaussian noise over $T$ diffusion steps, following a Markov chain \cite{ho2020denoising}. At each step $t$, noise $\epsilon \sim N(0,I)$ is added to the previous latent state $z_{t-1}$ according to a predefined noise schedule$\{\alpha_t, \sigma_t\}$:
\begin{equation}
z_t = \alpha_t z_0 + \sigma_t \epsilon.
\end{equation}
Conversely, the denoising process $p_{\theta}(z_{t-1}|z_t, t)$ iteratively removes noise over $T$ steps using a learned denoising network. Recently, v-prediction \cite{SalimansH22} has been proposed in the denoising process for enhanced numerical stability. This approach parameterizes the denoising direction as a velocity $v_t$, defined as a linear combination of the signal $z_0$ and noise $\epsilon$:
\begin{equation}
v_t = \alpha_t \epsilon - \sigma_t z_0.
\end{equation}
Unlike the original $\epsilon$-prediction, where the model predicts noise, the v-prediction model predicts the velocity. The training objective of the model is:
\begin{equation}
\mathcal{L}_{\theta}=\mathbb{E}_{z_0,t,\epsilon}[||v_t-v_{\theta}(z_t,t)||^2_2], 
\label{equ:loss}
\end{equation}
where $\theta$ is the model parameter.
During sampling at the step $t$, the denoised latent estimate $\hat{z}^t_{0}$ can be recovered from the predicted velocity using the following function:
\begin{equation}
\hat{z}^t_{0} = \alpha_t z_t - \sigma_t v_{\theta}(z_t,t).
\label{equ:x0}
\end{equation}

\vspace{-3mm}

\subsection{Hair Layering Modeling}\label{sec:hlm}
Our method begins by generating a 3D textured geometry $\mathcal{G}$ from the reference image $I_{\text{ref}}$. Following DrawingSpinUp \cite{zhou2024drawingspinup}, we first remove non-photorealistic contour lines from $I_{\text{ref}}$ to be a contour-free image \zj{$I^\text{nc}_{\text{ref}}$}. Next we use Wonder3D \cite{long2024wonder3d} to generate multi-view images ($I_{\text{front}}$, $I_{\text{right}}$, ..., $I_{\text{back}}$) and reconstruct a neural implicit signed distance field $\mathcal{I}$ from multi-view images. 
To address the unnatural deformation artifacts for characters with long hair (see Fig. \ref{fig:existing_problem} (a)-(b)), we have developed a Hair Layering Modeling (HLM) method to separate hair from the body. 
As shown in Fig. \ref{fig:hlm}, \zj{based on the manually-processed hair-body segmentation map provided by the user,} we first manually segment the front-view and right-view foreground maps to be the hair-body segmentation maps $S_{\text{front}}$ and $S_{\text{right}}$. Then we separate them into the hair segmentation maps ($S_{\text{front}}^{\text{hair}}$ and $S_{\text{right}}^{\text{hair}}$) and the body segmentation maps ($S_{\text{front}}^{\text{body}}$ and $S_{\text{right}}^{\text{body}}$). We also obtain an $S_{\text{back}}^{\text{hair}}$ by filling the internal region of $S_{\text{front}}^{\text{hair}}$ for the extraction of hair from the back of the head.
Finally, the implicit field separation is formulated as:
\begin{equation}
\begin{aligned}
\mathcal{S}_{\text{hair}} &= S_{\text{front}}^{\text{hair}} \cup \left( S_{\text{back}}^{\text{hair}} \cap S_{\text{right}}^{\text{hair}} \right), \\
\mathcal{S}_{\text{body}} &= S_{\text{front}}^{\text{body}} \cap S_{\text{right}}^{\text{body}}, \\
\mathcal{I}_{\text{hair}} &= \mathcal{I} \odot \mathcal{S}_{\text{hair}}, \\
\mathcal{I}_{\text{body}} &= \mathcal{I} \odot \mathcal{S}_{\text{body}},
\end{aligned}
\end{equation}
where $\odot$ represents the element-wise product. The individual geometries $\mathcal{G}_{\text{hair}}$ and $\mathcal{G}_{\text{body}}$ are reconstructed from their respective implicit fields \zj{$\mathcal{I}_{\text{hair}}$ and $\mathcal{I}_{\text{body}}$} via the Marching Cubes algorithm \cite{lorensen1998marching}, which are combined to form the final geometry $\mathcal{G}$. We then utilize the Mixamo \cite{mixamo} to create a rigged 3D character and retarget the target 3D motion onto it. 

\subsection{Guidance Sequence Rendering}\label{sec:recon}
\input{figures/fig_mask}
We generate three types of guidance sequences (i.e., pose $\{P_{1:N}\}$, \zj{SDI mask $\{M_{1:N}^{\text{SDI}}\}$} and \zj{coarse color $\{C_{1:N}\}$}) to provide geometric consistency for animation generation. We extract the pose sequence $\{P_{1:N}\}$ and reference pose $P_{\text{ref}}$ from the animated character using the OpenPose \cite{cao2019openpose} $18$-keypoint format, excluding hand and feet joints due to the abstract nature of hand-drawn characters.
As shown in Fig. \ref{fig:mask}, we allow users to provide optional \zj{SDI masks $M_{\text{front}}^{\text{SDI}}$ and $M_{\text{back}}^{\text{SDI}}$} to specify areas such as hair ends and skirts that require secondary dynamics enhancement (where a value of $1$ indicates the regions to be enhanced and a value of $0$ indicates the regions to be preserved). The \zj{SDI} masks corresponding to all the examples used in this paper are shown in Supplementary Materials Fig. 1.
We back-project the \zj{SDI} masks onto the geometry $\mathcal{G}$ to recolor each vertex as white (value of $1$) or black (value of $0$). Subsequently, we render the vertex colors as \zj{SDI} mask sequences \zj{$\{M_{1:N}^{\text{SDI}}\}$} to guide secondary dynamics injection. 
We render the animated 3D character back into the 2D image domain to generate a coarse color sequence $\{C_{1:N}\}$. We mask $\{C_{1:N}\}$ with \zj{$\{M_{1:N}^{\text{SDI}}\}$ to get a masked coarse color sequence $\{C^{\text{masked}}_{1:N}\}$ by \zj{$\{C^{\text{masked}}_{1:N}\}=\{C_{1:N}\} \cdot (1-M_{1:N}^{\text{SDI}}\})$}}. Leveraging our domain-adapted video diffusion model, we subsequently refine the preserved regions and redraw the masked regions using our Secondary Dynamics Injection strategy.

\subsection{Coarse Animation Refinement}\label{sec:refine}
Based on the three types of guidance sequences, our objective is to enhance the masked coarse color sequence by addressing two key aspects: insufficient appearance details and a lack of rich secondary motion. Ultimately, our domain-adapted model \zj{$v_{\theta}$} aims to generate a temporally coherent and realistic video sequence $\{\hat{I}_{1:N}\}$ from multiple conditions (i.e., $I_{\text{ref}}$, $P_{\text{ref}}$, $\{P_{1:N}\}$, $\{C^{\text{masked}}_{1:N}\}$ and \zj{$\{M_{1:N}^{\text{SDI}}\}$}).
We start with the state-of-the-art pose-controlled human animation method, UniAnimate \cite{wang2024unianimate}, which provides multiple natural motion priors. We then repurpose this generation task as an inpainting task tailored for the hand-drawn character domain. The rendered guidance sequences offer a stable, multi-view foundation during generation while significantly reducing the need for extensive tuning data.

\subsubsection{Model Architecture}\label{sec:model_arc}
\zj{As shown in Fig. \lz{\ref{fig:arch}}}, the main architecture of \zj{our domain-adapted model $v_{\theta}$} is similar to UniAnimate, with some minor modifications:
we add a coarse prior encoder $\mathscr{E}_{\text{coarse}}$ to embed the masked coarse color sequence $\{C^{\text{masked}}_{1:N}\}$ and the \zj{SDI} mask sequence \zj{$\{M_{1:N}^{\text{SDI}}\}$}. 
$\mathscr{E}_{\text{coarse}}$ is similar to the lightweight STC-encoder \cite{wang2023videocomposer}, consisting of \zj{two} 2D convolutional layers and a temporal Transformer layer to capture the spatial-temporal relations across all coarse frames. 
Specifically, $\{C^{\text{masked}}_{1:N}\}$ is concatenated with its corresponding \zj{$\{M_{1:N}^{\text{SDI}}\}$} along the channel axis and then fed into $\mathscr{E}_{\text{coarse}}$ to generate coarse prior embeddings. Such embeddings are subsequently concatenated with the input noise along the channel axis and then fed into the denoising UNet.
\input{figures/fig_framework}

\subsubsection{Model Tuning.}\label{sec:model_tune}
We prepare a hand-drawn animated video dataset (Section \ref{sec:dataset}) for model tuning to reduce the domain gap. 
Due to the significantly smaller size of our constructed dataset compared to the original training data for human dance videos (more than 10K), tuning the entire model on the limited motion variation and quantity poses a risk of catastrophic forgetting. Similar issues have been noted in ToonCrafter \cite{xing2024tooncrafter}. 
Following their strategy, we fine-tune the spatial layers to adapt the stylized appearance and freeze the temporal layers to maintain the real-world motion prior. Generally, the spatial layers primarily focus on appearance modeling, while the temporal layers aim to ensure motion coherence between frames.
Additionally, we train the coarse prior encoder $\mathscr{E}_{\text{coarse}}$ to enhance its ability to represent the coarse context and the intended refinement direction. \lz{The spatial layers and the coarse prior encoder are trained simultaneously using the general diffusion loss (Eq. \ref{equ:loss}).}

\subsection{Secondary Dynamics Injection}\label{sec:sec}
In skeletal animation, skinning weights are usually determined by the distance between vertices and the nearest bone. While skinning-based deformation effectively generates primary motion, it often fails to capture realistic secondary motion, such as cloth swaying or hair flowing. This issue is also evident in our synthesized dataset.
Although we freeze the temporal layers during training to preserve the natural motion priors inherent in real-world dynamics, the model's output after fine-tuning still more or less lacks the richness of secondary motion {(Section \ref{sec:ab_2})}.
Inspired by prior works \cite{kim2024diffusehigh, rout2024semantic, zhou2025light}, which leverage pre-trained diffusion models to guide generation for tasks such as super-resolution, image editing, and relighting, 
\lz{we propose Secondary Dynamics Injection to harness the real-world motion priors encoded in the native pre-trained UniAnimate, represented as $u_{\theta}$ (distinguished from our domain-adapted model $v_{\theta}$). This strategy guides the denoising process in our framework, as detailed in Supplementary Materials Algorithm 1.}

\subsubsection{Blending Latent Estimates}
\input{figures/fig_denoising}
To determine when and how to guide the denoising process, we begin by visualizing the frames decoded from the latent estimates \zj{$\{\hat{z}_{1:N,0}^{t, u_\theta}\}$}, computed according to the equation \ref{equ:x0}, at each denoising step $t$, as illustrated in Fig. \ref{fig:denoising}.
From this figure, we observed: 
(1) The initial denoising steps primarily establish the spatial structure of the frames, while the subsequent steps progressively refine the texture details. 
(2) In the steps associated with spatial distribution, the primary motion is generated first, with secondary motion details being gradually introduced in the later steps.
Based on these observations, we propose to blend the noise-free latent estimates \zj{$\{\hat{z}_{1:N,0}^{t, u_\theta}\}$} and \zj{$\{\hat{z}_{1:N,0}^{t, v_\theta}\}$} from the pre-trained model \zj{$u_{\theta}$} and our domain-adapted model \zj{$v_{\theta}$} separately, during the early denoising steps to introduce more dynamic motion, as the estimated $\hat{z}^t_{0}$ indicates the denoising direction in v-prediction \cite{SalimansH22} process.

\subsubsection{Reference Switching}
However, when guiding hand-drawn characters, the pre-trained model \zj{$u_{\theta}$} presents two key challenges previously mentioned (Fig. \ref{fig:existing_problem} (c) and (d)): (1) Difficulty in accurately modeling the motion of contour lines. (2) Inability to fully capture the semantics of the reference drawings.
To address these issues, during the initial denoising process, we utilize the contour-free reference image \zj{$I^\text{nc}_\text{ref}$} and leverage our domain-adapted model to produce initial latent estimates, which serve as a structurally aligned starting point that respects the reference style.
Specifically, we divide the generation process into three phases using two critical thresholds $\tau_2$ and $\tau_1$ ($T > \tau_2 > \tau_1$) expressed as percentages of the total timesteps ($\tau_2=\alpha \cdot T$ and $\tau_1=\beta \cdot T$ where $\alpha, \beta \in [0, 1]$). Here, $\tau_2$ and $\tau_1$ control the starting and ending step of secondary motion injection, respectively. 
First, during $[T, \tau_2]$, we employ only our model for denoising to ensure identity preservation and primary motion generation. 
Second, in $[\tau_2, \tau_1]$, we blend the two noise-free latent estimates with the downsampled masks \zj{$\{M^\text{SDI}_{1:N, \text{down}}\}$}. \lz{We use the $n$-th frame as an example to illustrate the latent fusion function}:
\zj{
\begin{equation}
 \hat{z}_{n, 0}^{t, \text{blend}} = (1-M^\text{SDI}_{n, \text{down}}) \cdot \hat{z}_{n,0}^{t, v_\theta} + M^\text{SDI}_{n, \text{down}} \cdot \hat{z}_{n,0}^{t, u_\theta}.
\end{equation}
}

The combination of latent estimates will provide a new optimization direction. While replacing the contour-free reference $I^\text{nc}_\text{ref}$ with the original reference $I_{\text{ref}}$ during later denoising steps $[\tau_1, \tau_0]$ introduces contour structures, the results often fail to faithfully preserve the reference's contours (as analyzed in Section \ref{sec:ab_3}). 
Thus, in the third phase, we first augment the masked regions in $\{C^\text{masked}_{1:N}\}$ with the estimated video \lz{$D(\{\hat{z}_{1:N, 0}^{\tau_1,\text{blend}}\})$, }
\lz{where $D(\cdot)$ is the VAE decoder}, to obtain the inpainted coarse frames $\{C^\text{inpainted}_{1:N}\}$ via Poisson Blending \cite{perez2003poisson}.
\zj{As illustrated in Fig. \ref{fig:poisson}, compared with directly concatenating $C^\text{masked}_{n}$ with the estimated video \lz{$D(\{\hat{z}_{n, 0}^{\tau_1,\text{blend}}\})$}, blending them using Poisson Blending achieves more natural stitching.
\input{figures/fig_poisson}
}

\subsubsection{Re-denoising}
Constrained with the updated coarse inputs with rich dynamics and the reference image $I_{\text{ref}}$ with contours, we employ our domain-adapted model $v_{\theta}$ to re-denoise the initial noise $\{z_{1:N, T}\}$ from scratch.
Since only early denoising steps influence the motion distribution, \zj{when $\tau_1$ < 0.7, the impact on motion distribution remains minimal; however, this reduction in $\tau_1$ leads to an increase in processing time.}

%% file: figures/fig_hlm.tex
\begin{figure*}[t]
    \centering
    \includegraphics[width=0.85\textwidth]{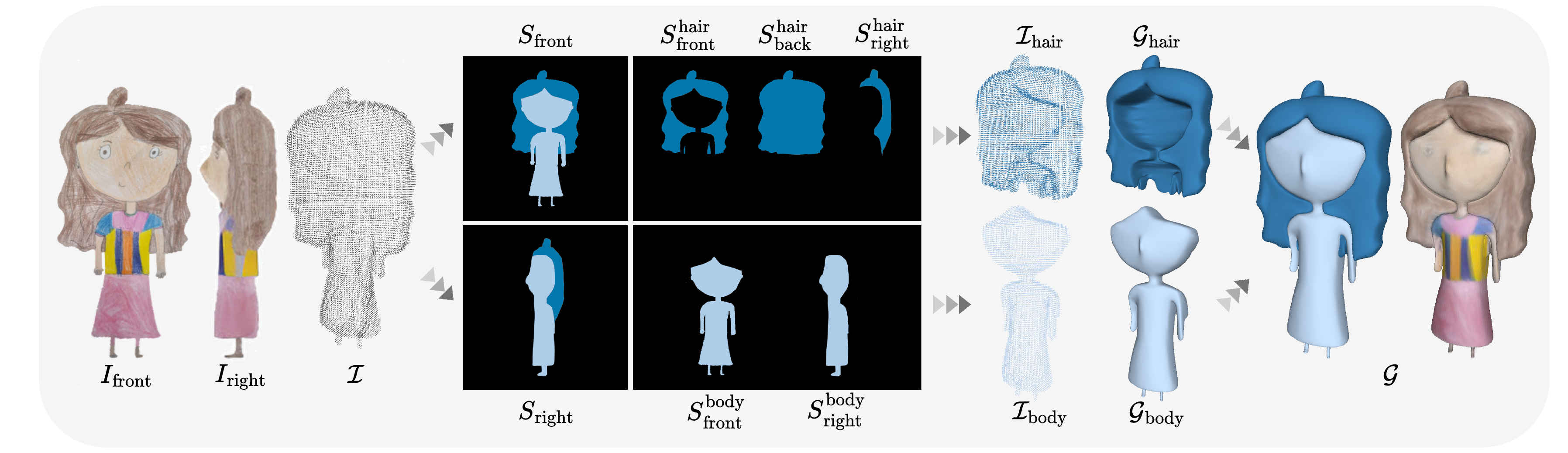}
    \vspace{-3mm}
    \caption{An illustration of hair layering modeling.}
    \vspace{-3mm}
    \label{fig:hlm}
\end{figure*}

%% file: figures/fig_mask.tex
\begin{figure}[h]
    \centering
    \includegraphics[width=0.48\textwidth]{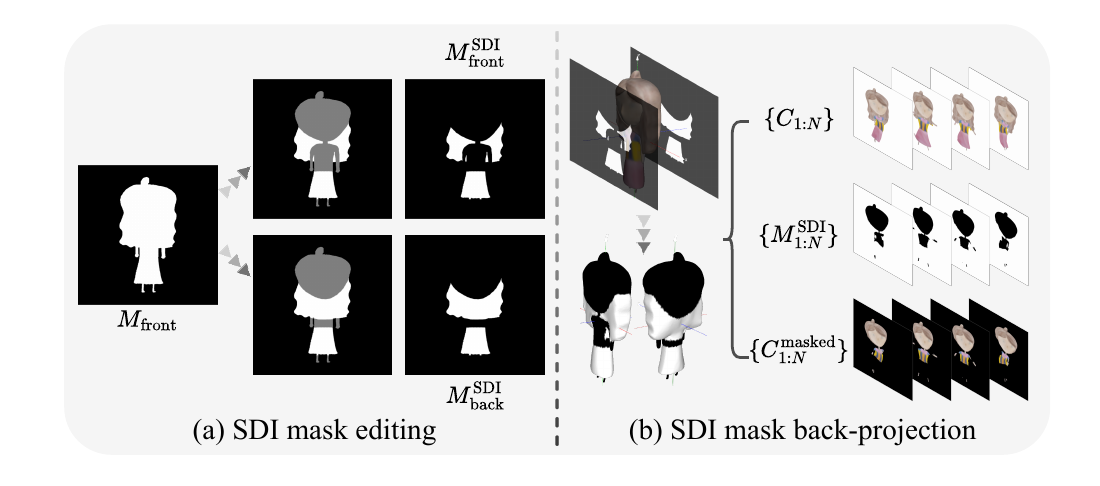}
    \vspace{-6mm}
    \caption{An illustration of 
    how we obtain the SDI mask sequence.}
    \vspace{-3mm}
    \label{fig:mask}
\end{figure}

%% file: figures/fig_framework.tex
\begin{figure}[h]
    \centering
    \includegraphics[width=0.48\textwidth]{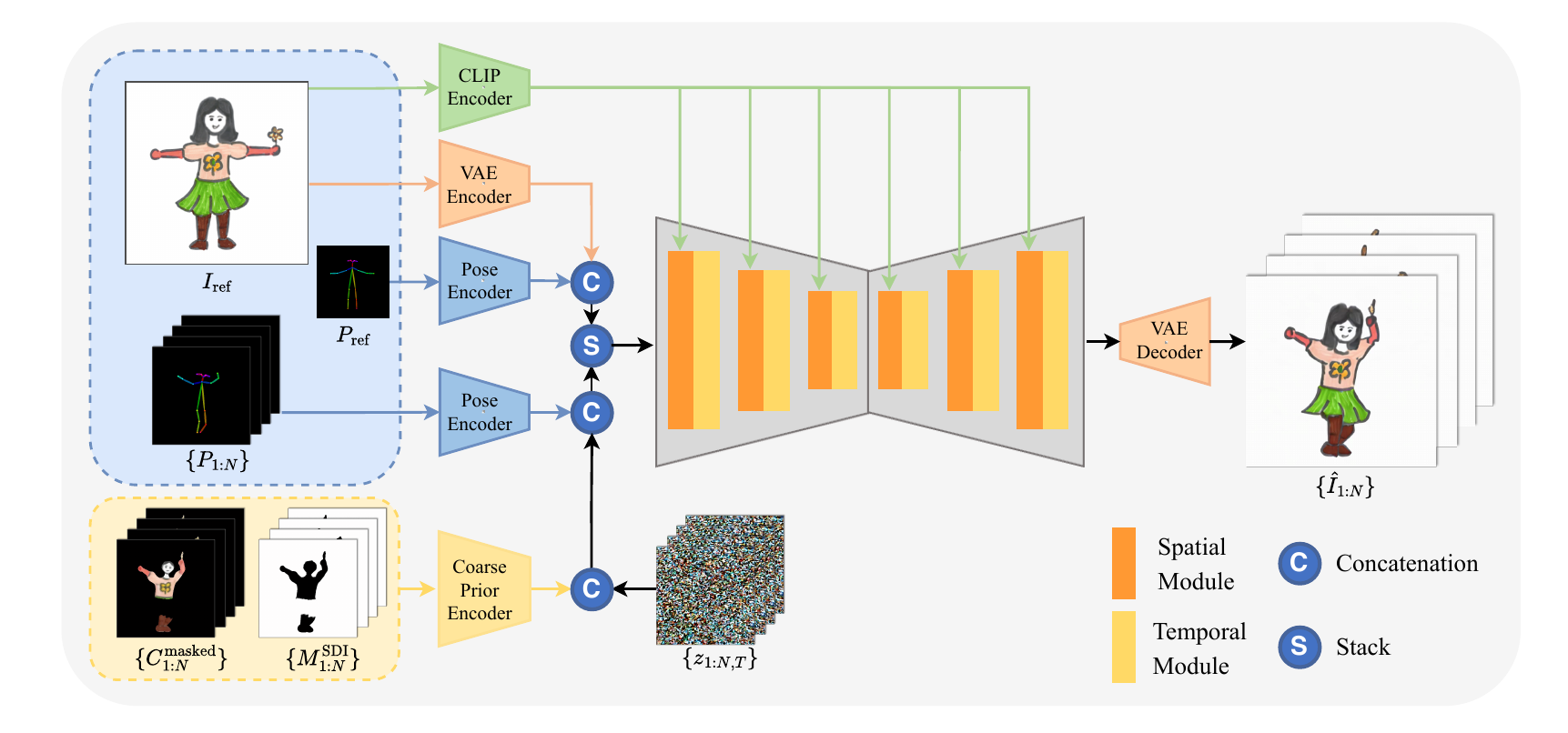}
    \vspace{-6mm}
     \caption{An illustration of our domain-adapted diffusion model $v_{\theta}$.}
    \vspace{-3mm}
    \label{fig:arch}
\end{figure}

%% file: figures/fig_denoising.tex
\begin{figure*}[h]
    \centering
    \includegraphics[width=1\textwidth]{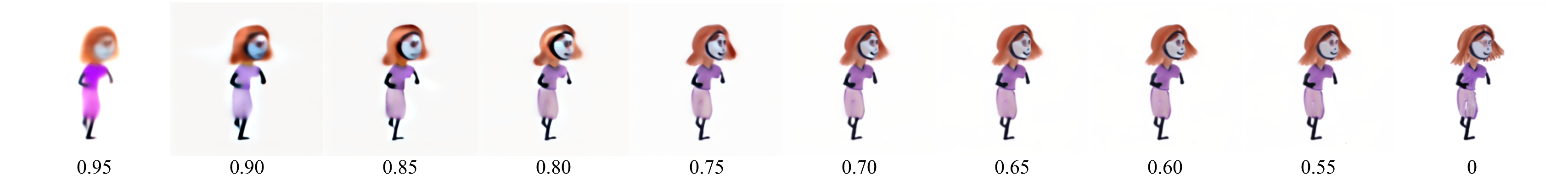}
    \vspace{-6mm}
    \caption{
    Gradually decreasing t-step denoised latent estimation results. The numbers below the images represent the percentage of the denosing process.}
    \vspace{-3mm}
    \label{fig:denoising}
\end{figure*}

%% file: figures/fig_poisson.tex
\begin{figure}[h]
    \centering
    \includegraphics[width=0.48\textwidth]{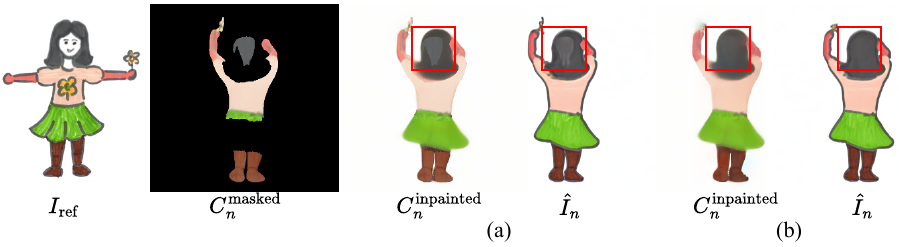}
    \vspace{-6mm}
    \caption{An example illustrating Poisson Blending for inpainting the n-th masked coarse color image. (a) shows the inpainting of the n-th masked coarse color image $C^\text{masked}_{n}$ by directly concatenating it with the estimated video $D(\{\hat{z}_{n, 0}^{\tau_1,\text{blend}}\})$. (b) demonstrates the inpainting process by mixing them using Poisson Blending.
 }
    \vspace{-3mm}
    \label{fig:poisson}
\end{figure}

%% file: files/evaluation.tex
\section{Experiments}
\subsection{Dataset Construction} \label{sec:dataset}
Due to the scarcity of hand-drawn animated video data that features plausible secondary motion, we created a small-scale primary animation dataset as a compromise. 
This dataset is used to help the model refine texture details and contour lines, reducing the gap between real-human and hand-drawn characters.
We collect $174$ high-quality character drawing images as training and evaluation data from two existing datasets: the Amateur Drawings Dataset \cite{smith2023method} and the SketchAnim Dataset \cite{rai2024sketchanim}. 
Since Mixamo's auto-rigging tool may not function properly if the character is significantly asymmetric or posed, before rigging, we applied rotation or local deformation to the severely asymmetric examples in these images to prevent rigging failures.
Each character is assigned 1-2 motions selected from the Mixamo Animation Dataset, varying in length from a few dozen to several hundred frames, encompassing both simple and complex motions. To improve multi-view consistency, the training set also includes a 60-frame full rotation of the character in a rest pose.
We use the stylized animation videos generated by DrawingSpinUp \cite{zhou2024drawingspinup} as ground truth to fine-tune our diffusion models. 
Ultimately, we obtained $428$ high-quality drawing animation video clips, which were randomly divided into two sets: the training set contains $359$ clips covering $124$ different characters, while the evaluation set includes $69$ clips with $50$ different characters.

\zj{
\subsection{Implementation Details} \label{sec:imp}
The experiments were conducted using a single 80G NVIDIA A100 GPU. We optimized our diffusion model $v_{\theta}$ using an AdamW optimizer over 40k steps, with a learning rate of 2e-5 and a batch size of 4. For training videos, we sampled 16 frames from each video and randomly cropped a $768 \times 512$ region.  In each iteration, we randomly decided whether to apply a mask, helping the model to refine textures in unmasked regions. We enhanced mask diversity by incorporating random masks and square masks, adapted from ProPainter \cite{zhou2023propainter}.
During inference, we adopted the DDIM sampler \cite{song2020denoising} with 20 steps. Based on experimental results, we recommend setting the hyperparameters: $\alpha$ in the range $[0.7, 0.95]$ and $\beta$ in the range $[0.5, 0.7]$.
For long video generation, we adopt the sliding window strategy proposed in \cite{zhu2024champ, xu2024magicanimate}, synthesizing videos segment by segment with temporally overlapping frames. For the input conditions of overlapping frames, we leverage inpainted coarse frames $\{C^{\text{inpainted}}_{1:N}\}$ generated from previous segments to guide the generation of subsequent segments.
Given a character drawing, it takes 3-5 minutes to generate a rigged 3D character (done only once for retargeting various animations). Our model generates a 32-frame video at a resolution of $768 \times 512$ in 45 seconds on a single A100 GPU without the Secondary Dynamics Injection strategy. When SDI is introduced, the total time increases to 72, 81, and 90 seconds for $\beta$ values of 0.7, 0.6, and 0.5, respectively.
}

\subsection{Qualitative Results} \label{sec:com}
We evaluate our method against state-of-the-art animation methods, categorized into two paradigms: 
(1) traditional skeletal animation methods: Smith et al. \shortcite{smith2023method} and DrawingSpinUp \cite{zhou2024drawingspinup}; 
(2) diffusion-based methods: AnimateAnyone \cite{hu2024animate}, MikuDance \cite{zhang2024mikudance}, and UniAnimate \cite{wang2024unianimate}. 
Additionally, we implemented UniAnimate* by fine-tuning it on our synthesized dataset. 
\zj{For Smith et al. \shortcite{smith2023method} and DrawingSpinup \cite{zhou2024drawingspinup}, we utilize their official implementations. 
For UniAnimate \cite{wang2024unianimate}, we also employ their official implementation.
Additionally, we fine-tune UniAnimate on our synthesized dataset, following a similar tuning approach as described in Section \ref{sec:imp}, referred to as UniAnimate*.
For AnimateAnyone \cite{hu2024animate}, we generate results using the publicly available reproduced code \cite{mooreanimateanyone}, as the official implementation was not accessible. 
For MikuDance \cite{zhang2024mikudance}, we use the official code but disable scene motion tracking to ensure a fair comparison, given that our setup assumes a clean white background.}

\input{figures/fig_res_0}
\emph{Ours vs Skeletal Animation Methods.} Fig. \ref{fig:res_0} compares our method with two traditional skeletal animation techniques. Our observations are as follows.
Both DrawingSpinUp and our method, supported by 3D modeling, yield more plausible 3D-aware animations that faithfully represent the target motions than Smith et al.
Since skeletal animation can only generate primary motion, Smith et al. and DrawingSpinUp struggle to produce natural secondary movements. In contrast, our method effectively captures subtle dynamics, such as the swinging motion of the little girl's ponytails as she jumps (Row 1).
In the results of Smith et al. and DrawingSpinUp, unnatural deformation artifacts occur regardless of whether a 2D or 3D mesh is used, as seen in the entanglement between the hair and shoulders of the characters (Row 2-3). In contrast, our method animates these challenging long-hair cases more naturally.

\input{figures/fig_res_1}
\emph{Ours vs Diffusion-Based Methods.} Fig. \ref{fig:res_1} compares our method with several diffusion-based animation approaches. Our method demonstrates robust generalization across diverse character styles, from humanoid designs to highly stylized drawings, while faithfully preserving identity and delivering smooth, natural animations.
Diffusion-based methods show limited ability for humanoid characters. They often struggle to maintain view-consistent contour lines due to the absence of such features in their training data (Row 2).
Fine-tuning with our domain-specific data successfully achieves stylized appearance adaptation, as demonstrated in Fig. \ref{fig:res_1} (Column \emph{UniAnimate} and \emph{UniAnimate*}).
However, these methods frequently misinterpret the 2D skeleton, resulting in unnatural or erroneous animations when handling complex 3D motions (Rows 1 and 3). 
We hypothesize that the inherent limitations of 2D skeletons, particularly their inability to adequately represent occlusions and depth variations in intricate 3D movements, are the primary cause of these errors.
To address this issue, our method leverages coarse input to enrich primary motion understanding, enabling more accurate and realistic animations in challenging scenarios.
When applied to characters with styles that deviate significantly from typical human appearances, methods like UniAnimate and AnimateAnyone often misclassify parts, or even the entirety, of the character as background elements (Rows 1 and 3). This results in incomplete or entirely static animations, highlighting their inability to adapt to nonstandard character designs.
Although MikuDance benefits from extensive training on large-scale anime datasets, it still struggles to generalize effectively to hand-drawn characters. Its reliance on specific data distributions limits adaptability to characters with unique stylistic elements or unconventional features.

\subsection{Quantitative Results}
We conducted a quantitative evaluation to assess the texture quality of the animation results. We measure LPIPS \cite{zhang2018unreasonable} for texture consistency, FID \cite{heusel2017gans} for overall distribution similarity, and CLIP \cite{radford2021learning} similarity for assessing semantic alignment. These metrics are calculated between the generated frames and the reference images.
Considering that UniAnimate, AnimateAnyone, and MikuDance fail to accurately animate the reference image with the 3D poses, they may directly maintain some texture in the reference images, as shown in Fig. \ref{fig:res_1}. This may lead to biased results. Therefore, we omit comparisons with these methods, focusing instead on DrawingSpinUp and UniAnimate*, which are better at achieving accurate pose variation. 
Table \ref{tab:qua1} presents the quantitative results, and our method outperforms all compared approaches, demonstrating its ability to faithfully restore the texture of stylized hand-drawn characters across various 3D poses.

\begin{table}[htbp]
\begin{center}
\scalebox{0.9}{
\begin{tabular}{ccccc}
    \hline
	\multicolumn{1}{c}{\textbf{Method}}
        & \multicolumn{1}{c}{\textbf{LPIPS $\downarrow$}}
        & \multicolumn{1}{c}{\textbf{FID $\downarrow$}}
        & \multicolumn{1}{c}{\textbf{CLIP $\uparrow$}}\\
	\hline
    UniAnimate* & 0.1792 & 158.1467 & 0.8964 \\
    DrawingSpinUp  & 0.1734 & 157.7452 & 0.8880  \\
    \hline
    Ours  & \textbf{0.1733} & \textbf{152.9022} & \textbf{0.9030} \\
    \hline
\end{tabular}
}
\end{center}
\caption{Quantitative comparisons with two pose animation methods.}
\vspace{-6mm}
\label{tab:qua1}
\end{table}

\subsection{Ablation Study}

\subsubsection{Ablation on HLM}

We conducted an ablation study to verify the effectiveness of our Hair Layering Modeling (HLM). As shown in Fig. \ref{fig:ab0}, the unnatural deformation in the face and armpits caused by the hair adhering to the shoulders adversely affects subsequent animation generation, degrading the final animation results.

\subsubsection{Ablation on Training Components of $v_{\theta}$}

To validate the effectiveness of our domain-adapted diffusion model $v_{\theta}$, we conducted an ablation study to demonstrate the contributions of two key components, the coarse prior encoder $\mathscr{E}_{\text{coarse}}$ and the choice of spatial layer tuning (SLT). 
We compare the following configurations: 
(I) Ablating the coarse prior encoder $\mathscr{E}_{\text{coarse}}$;
(II) Replacing spatial layer tuning (SLT) with temporal layer tuning (TLT);
(III) Our method. 
As shown in Fig. \ref{fig:ab1}, we can see that changing the key components significantly degrades performance, especially in 3D motion understanding.
Although ablating $\mathscr{E}_{\text{coarse}}$ achieves stylized appearance adaptation, it fails to produce plausible results for complex poses involving occlusions or intricate 3D spatial relationships. This leads to limb confusion (e.g., left-right arm swaps) and the erroneous generation of occluded limbs that should be hidden. 
As mentioned in Section \ref{sec:com}, the limitation of the 2D pose input is its inability to explicitly encode 3D spatial priors, such as depth ordering and occlusion cues, leading to geometric inconsistencies. 
$\mathscr{E}_{\text{coarse}}$ addresses this by introducing dense, structured 3D guidance, which disambiguates limb positions and enforces physically plausible occlusion reasoning. 
Replacing SLT with TLT achieves partial adaptation but fails to establish correct relationships between generated content and the reference image for masked regions. This is evidenced by artifacts such as hat discontinuities (Row 1) and erroneous texture propagation (e.g., using trouser patterns to synthesize hair) (Row 2). These results highlight that spatial layers play a more critical role in appearance modeling and should be prioritized during fine-tuning for robust adaptation.

\zj{
\subsubsection{Ablation on SDI} \label{sec:ab_2}
To evaluate the effectiveness of our Secondary Dynamics Injection (SDI) strategy, we performed an ablation study that compares scenarios with and without SDI. As shown in Fig. \ref{fig:ab2}, compared with only using the domain-adapted model without SDI, our SDI strategy effectively generates secondary dynamics, extending beyond the primary dynamics.

\input{figures/fig_ab0}
\input{figures/fig_ab1}
\input{figures/fig_ab2}

\subsubsection{Ablation on Key Components of SDI} \label{sec:ab_3}
To further evaluate the modular design of our Secondary Dynamics Injection (SDI) strategy, we conducted an ablation study to demonstrate the contributions of sub-modules: blending (Bld) with the pre-trained model $u_{\theta}$, reference switching (RS) between the reference image $I_\text{ref}$ and the contour-free reference image \zj{$I^\text{nc}_\text{ref}$}, and re-denoising (RD). 
We compare the following configurations, with all comparisons using our domain-adapted model $v_{\theta}$ during denoising steps $[T, \tau_2]$: 
(I) Directly using $v_{\theta}$ conditioned on $I_\text{ref}$ during steps $[\tau_2,0]$ (\emph{W/o SDI}); 
(II) Blending the latent estimates from $v_{\theta}$ and $u_{\theta}$ conditioned on $I_\text{ref}$ during steps $[\tau_2,0]$ (\emph{Bld}); 
(III) Blending the latent estimates from $v_{\theta}$ and $u_{\theta}$, initially conditioned on $I^\text{nc}_\text{ref}$ during steps $[\tau_2, \tau_1]$ and then switching to $I_\text{ref}$ during $[\tau_1, 0]$ (\emph{Bld+RS}); 
(IV) Blending the latent estimates from $v_{\theta}$ and $u_{\theta}$, initially conditioned on $I^\text{nc}_\text{ref}$ during steps $[\tau_2, \tau_1]$ and then re-denoising from initial noise during steps $[T, 0]$ with $I_\text{ref}$ (\emph{Bld+RS+RD}).

As shown in Fig. \ref{fig:ab3}, only using the domain-adapted model (Column \emph{W/o SDI}) captures primary motion but fails to generate secondary dynamics. Blending with the pretrained model \zj{$u_\theta$} using contour references (Column \emph{Bld}) often misplaces contour patterns into internal regions, primarily because of the incorrect texture propagation of the pretrained model \zj{$u_\theta$}.
Initial use of contour-free references (Column \emph{Bld+RS}) mitigates these artifacts. However, late-stage introduction of contour references at $[\tau_1, 0]$ leads to visible discrepancies in generated contours and inconsistent contour style preservation. Thus, our final solution (Column \emph{Bld+RS+RD}) addresses these limitations through intermediate coarse video estimation and blending to ensure faithful motion injection and consistent style preservation.
\input{figures/fig_ab3}
}
\input{figures/fig_ab4_1}
\input{figures/fig_ab4_2}

\subsubsection{Ablation on Key Parameters of SDI} \label{sec:ab_4}
To evaluate the effectiveness of our Secondary Dynamics Injection (SDI) strategy, we further examined the effects of varying the two key parameters, $\alpha$ and $\beta$, \lz{which define the thresholds $\tau_2$ and $\tau_1$, determining the transitions between different phases of the SDI process.}
We first vary $\alpha$ at values between 0.70 and 0.95 while keeping $\beta$ fixed at 0.60. Then we set $\alpha=0.95$ and vary $\beta$ at values of 0.50, 0.60, and 0.70. 
As shown in Fig. \ref{fig:ab4_1}, $\alpha$ significantly influences the initiation of the injection from the pretrained model \zj{$u_\theta$}. 
When $\alpha$ is at relatively low values (e.g., 0.7), the results are similar to those without SDI. As $\alpha$ increases, the motion intensity becomes more pronounced, until at very high values (e.g., 0.95), some artifacts begin to appear. Therefore, there is a reasonable range for the values of $\alpha$, enabling users to achieve the desired effect by adjusting $\alpha$. 
Fig. \ref{fig:ab4_2} illustrates the inpainted coarse color images $\{C^\text{inpainted}_{1:N}\}$ along with the corresponding results $\{\hat{I}_{1:N}\}$. It is observed that the results do not exhibit significant variations when $\beta$ changes. This also demonstrates the robustness of our coarse prior encoder. From the perspective of minimizing runtime, we select 0.6 or 0.7 as an appropriate value for $\beta$.

%% file: figures/fig_res_0.tex
\begin{figure*}[h]
    \centering
    \includegraphics[width=1\textwidth]{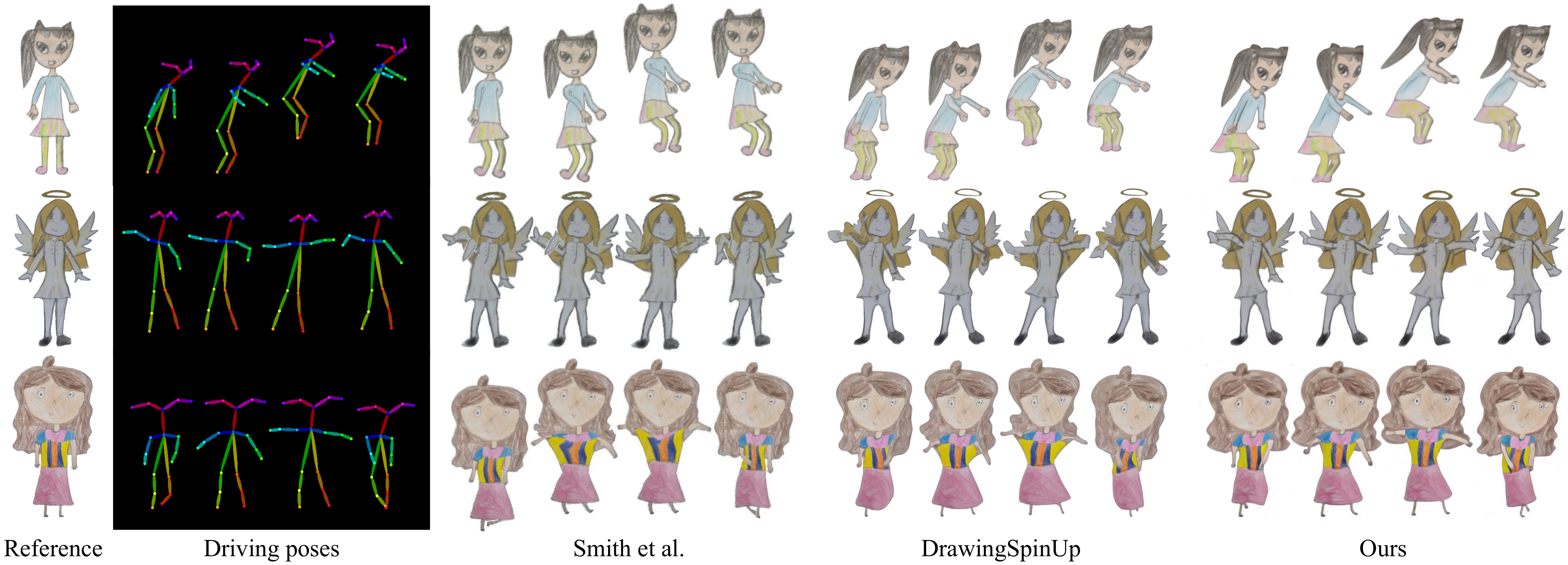}
    \vspace{-6mm}
    \caption{Visual comparisons {with two traditional skeletal animation methods.}}
    \label{fig:res_0}
\end{figure*}

%% file: figures/fig_res_1.tex
\begin{figure*}[h]
    \centering
    \includegraphics[width=1\textwidth]{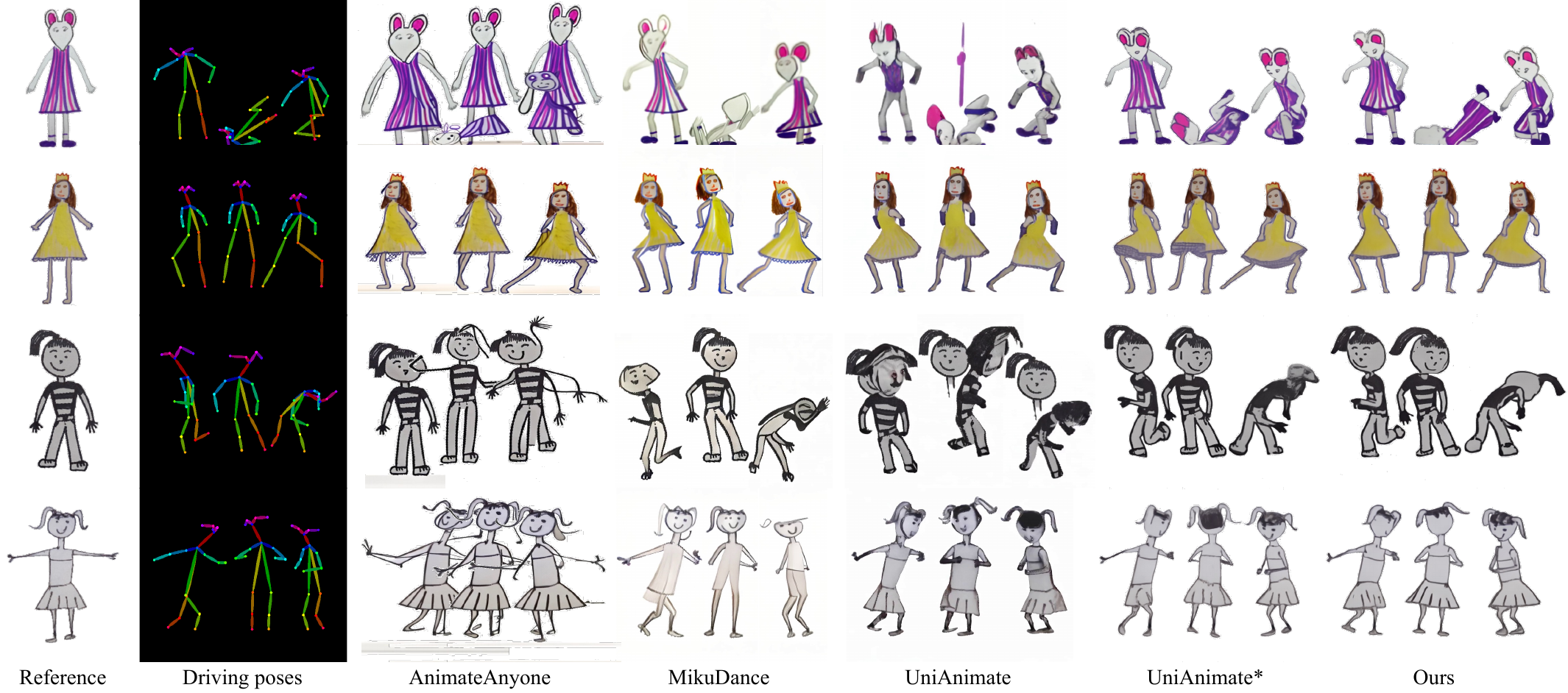}
    \vspace{-6mm}
    \caption{Visual comparisons 
    {with four diffusion-based animation methods.}}
    \label{fig:res_1}
\end{figure*}

%% file: figures/fig_ab0.tex
\begin{figure}[h]
    \centering
    \includegraphics[width=0.48\textwidth]{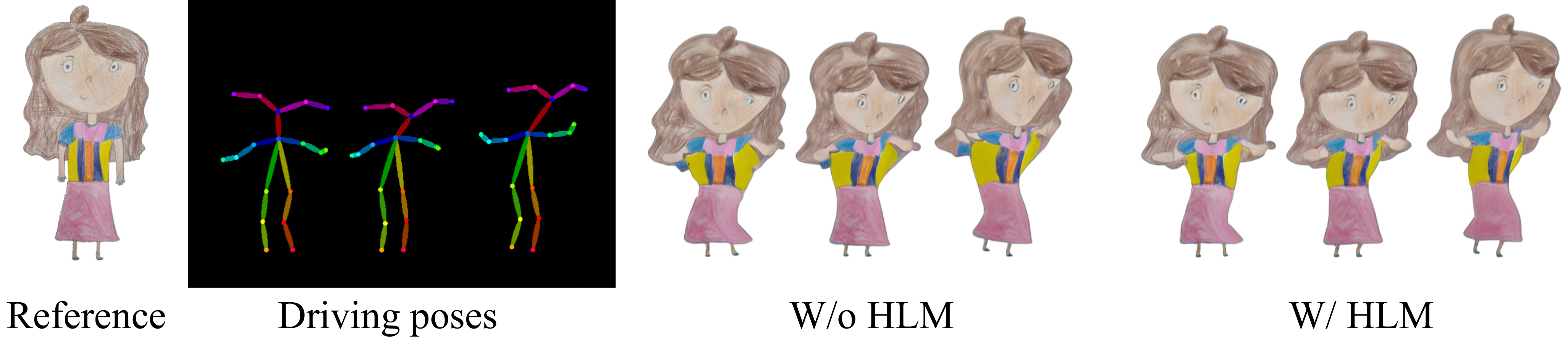}
    \vspace{-6mm}
    \caption{Ablation on our hair layering modeling (HLM) method.}
    \vspace{-3mm}
    \label{fig:ab0}
\end{figure}

%% file: figures/fig_ab1.tex
\begin{figure}[h]
    \centering
    \includegraphics[width=0.48\textwidth]{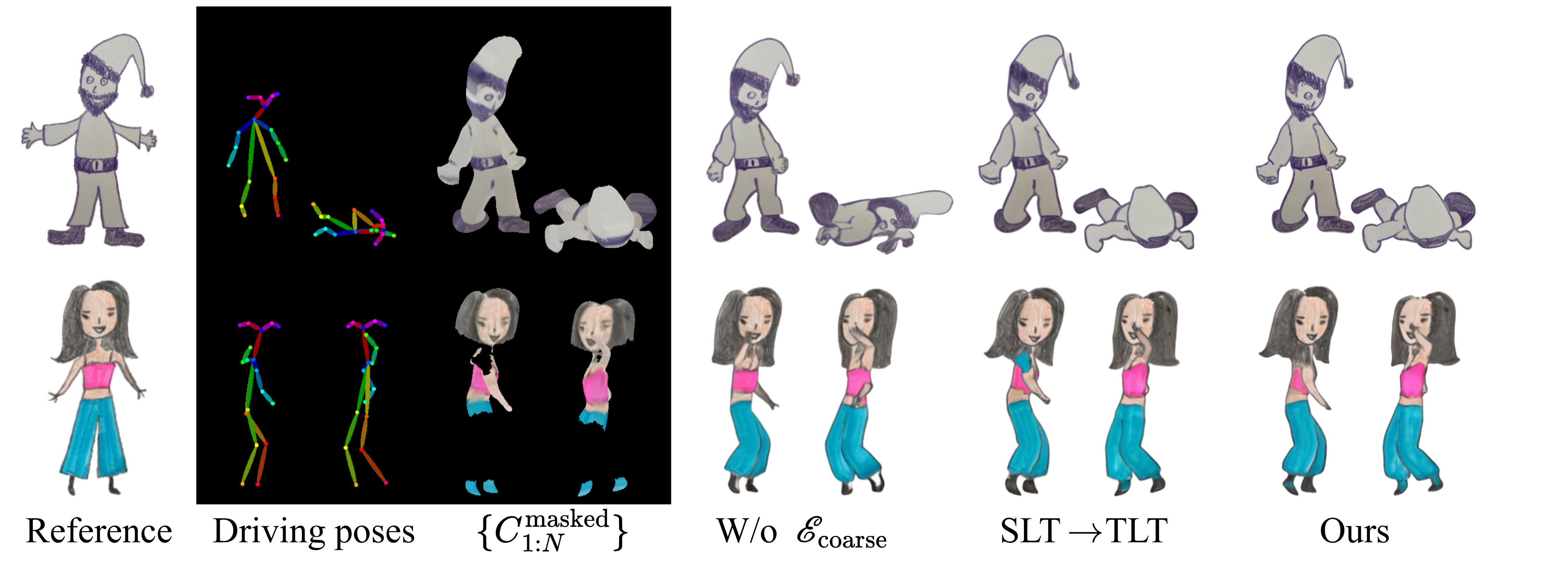}
    \vspace{-6mm}
    \caption{Ablations on training  components of $v_\theta$.}
    \vspace{-3mm}
    \label{fig:ab1}
\end{figure}

%% file: figures/fig_ab2.tex
\begin{figure}[h]
    \centering
    \includegraphics[width=0.48\textwidth]{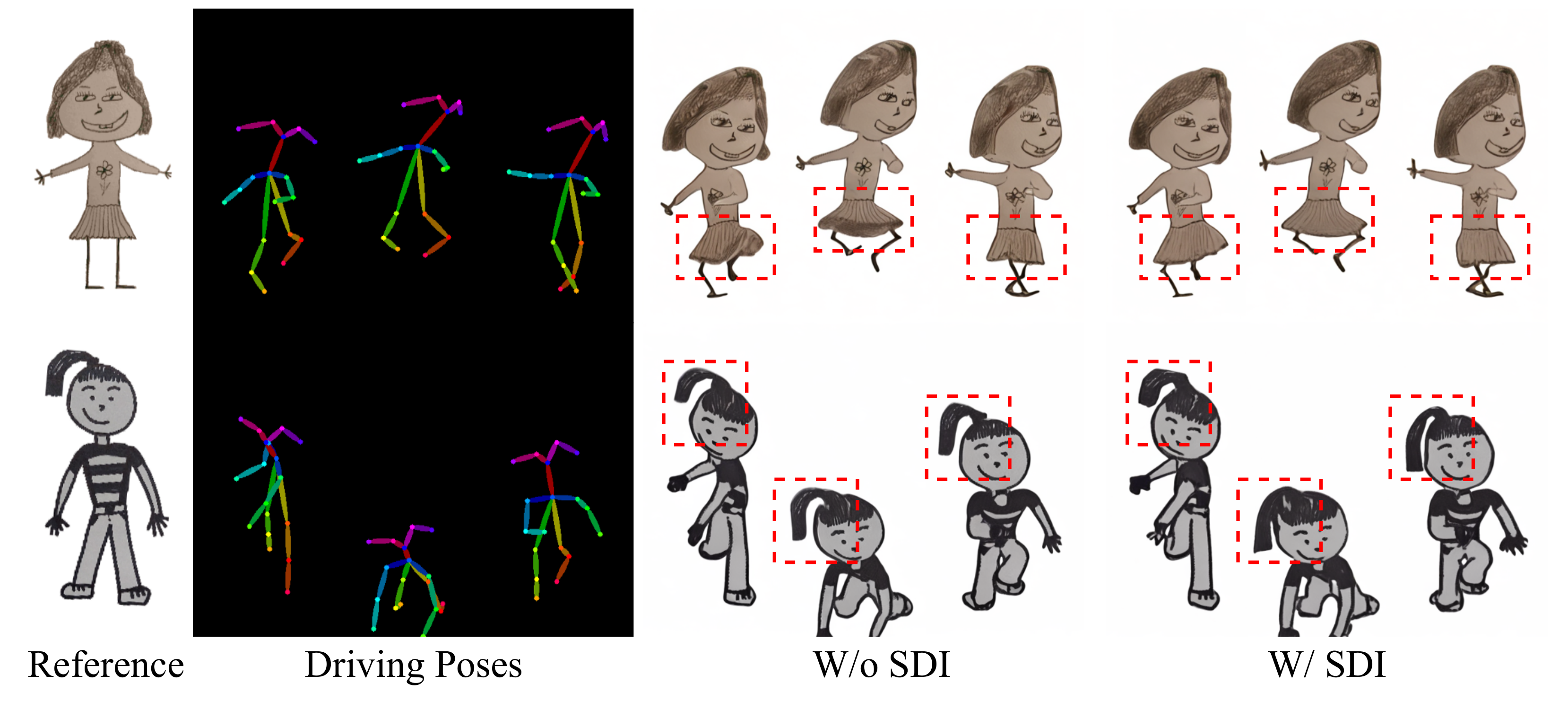}
    \vspace{-6mm}
    \caption{Ablations on our Secondary Dynamics Injection (SDI) strategy.}
    \vspace{-3mm}
    \label{fig:ab2}
\end{figure}

%% file: figures/fig_ab3.tex
\begin{figure}[h]
    \centering
    \includegraphics[width=0.48\textwidth]{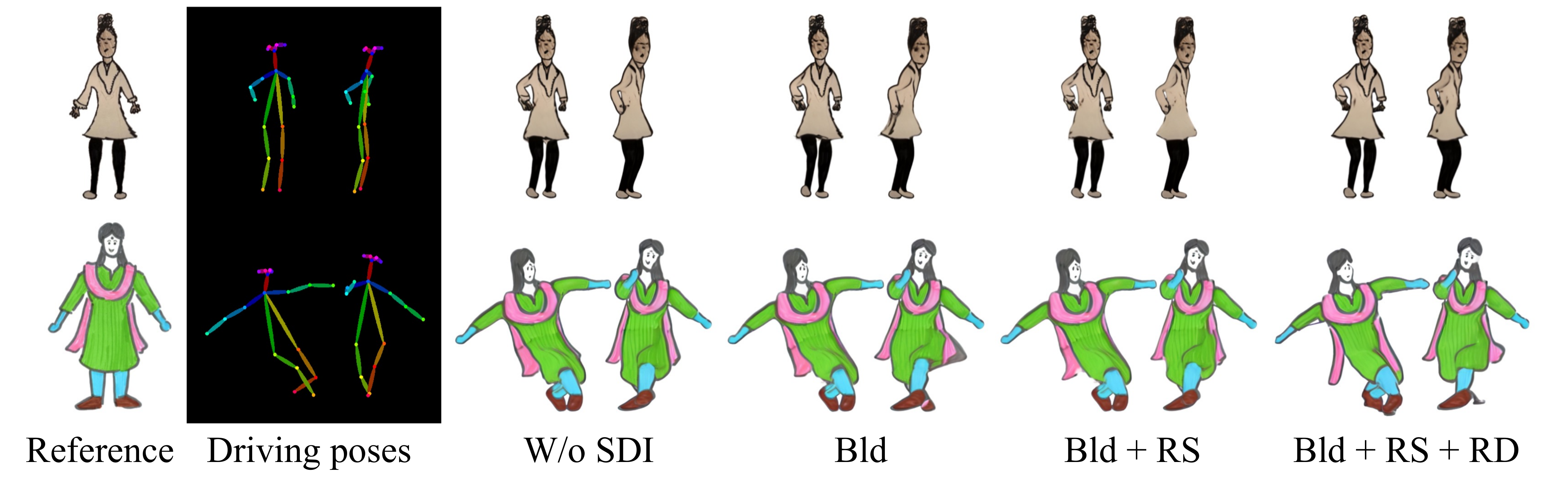}
    \vspace{-6mm}
    \caption{Ablations on three key components of our Secondary Dynamics Injection (SDI) strategy. The following abbreviations are used: Bld for blending, RS for reference switching, and RD for re-denoising.}
    \vspace{-3mm}
    \label{fig:ab3}
\end{figure}

%% file: figures/fig_ab4_1.tex
\begin{figure}[h]
    \centering
    \includegraphics[width=0.48\textwidth]{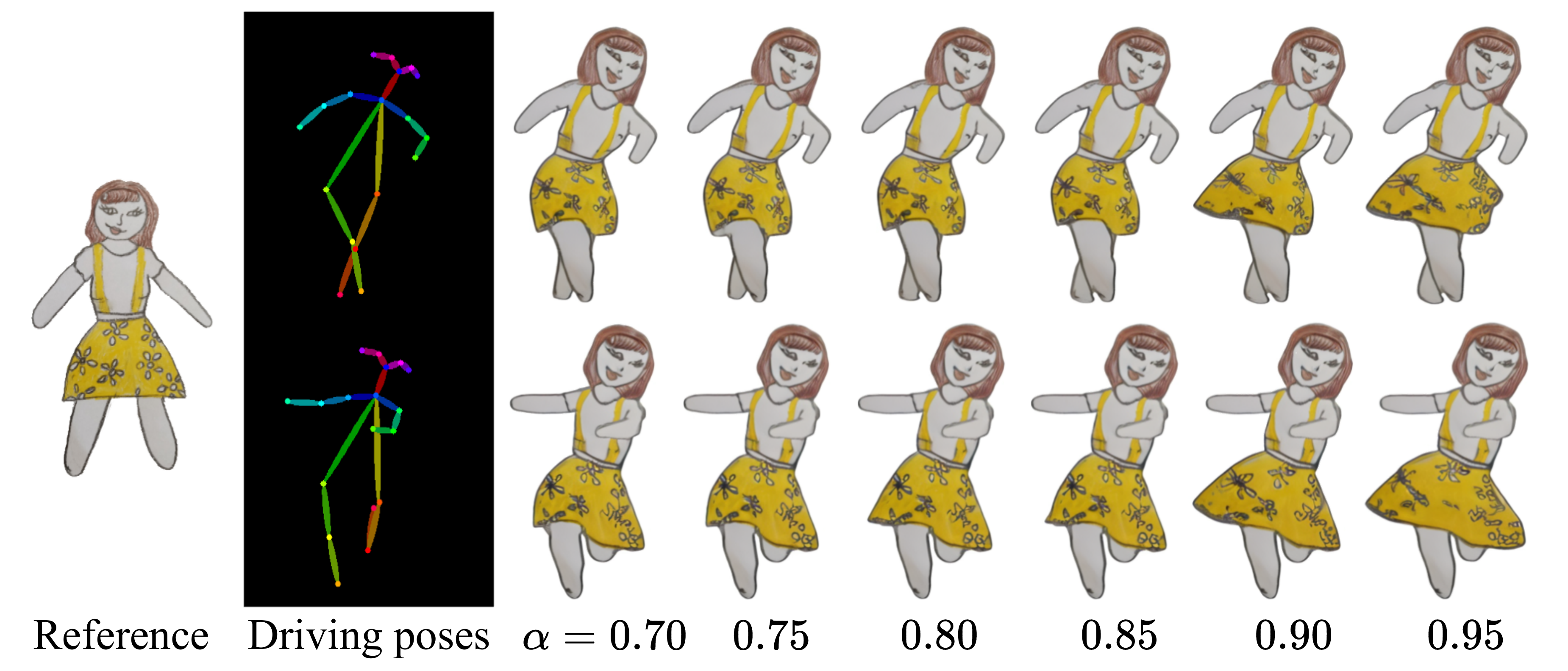}
    \vspace{-6mm}
    \caption{The impact of different values of $\alpha$ between $0.70$ and $0.95$ on the secondary dynamics of the animation results, with $\beta=0.60$.}
    \vspace{-3mm}
    \label{fig:ab4_1}
\end{figure}

%% file: figures/fig_ab4_2.tex
\begin{figure}[h]
    \centering
    \includegraphics[width=0.48\textwidth]{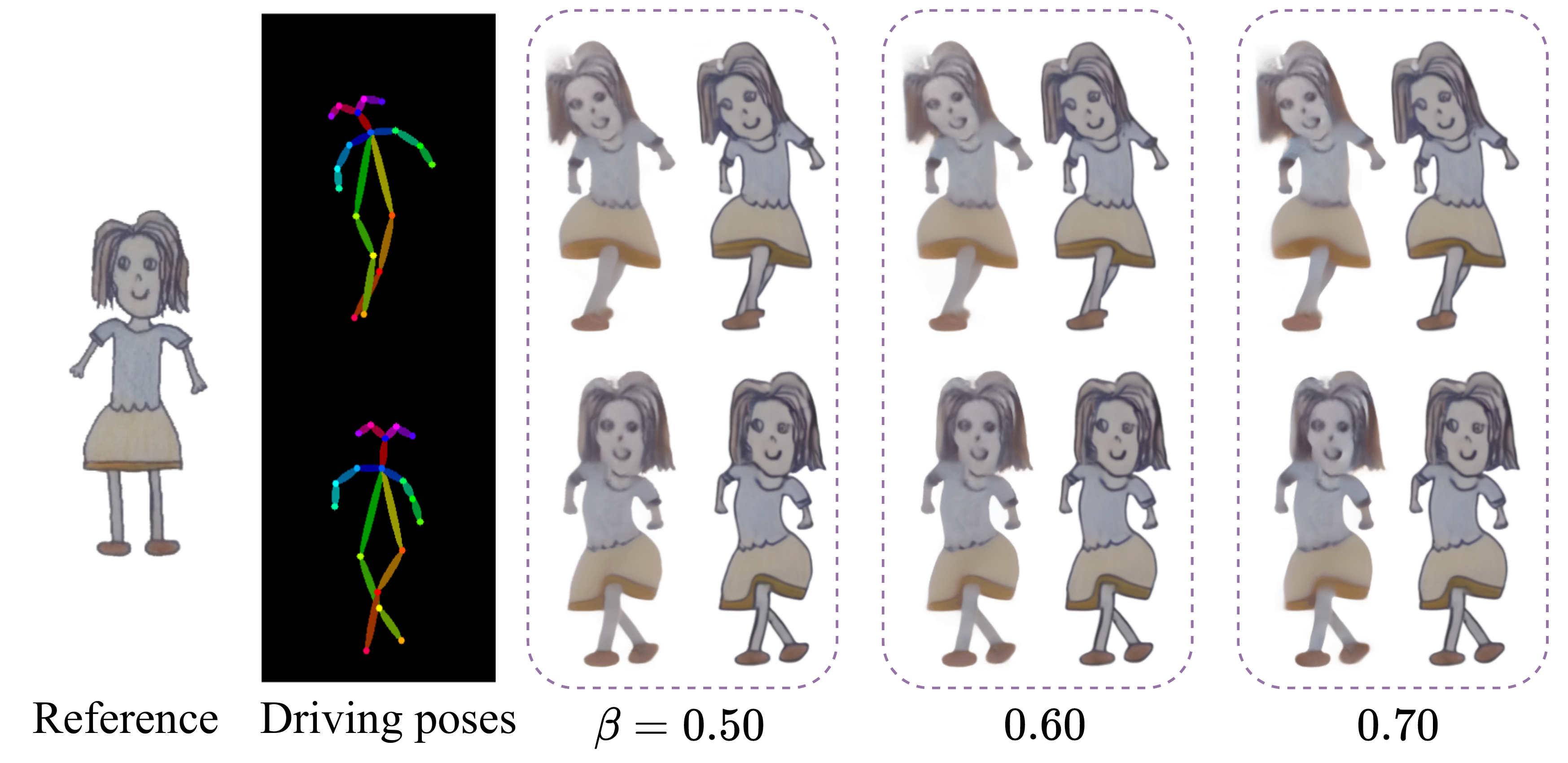}
    \vspace{-6mm}
    \caption{The impact of different values of $\beta$ ($0.50$, $0.60$, $0.70$) on the secondary dynamics of the animation results, with $\alpha=0.95$.
    }
    \vspace{-3mm}
    \label{fig:ab4_2}
\end{figure}

%% file: files/application.tex
\section{Application}
\input{figures/fig_app}
Our system supports user-friendly editing, allowing users to change characters' appearances with just a few strokes. Users can locally erase and repaint on the reference image, and the system updates the SDI mask sequence accordingly. As shown in Fig. \ref{fig:app}, users can edit character patterns and designs, as well as modify shape features like hairstyle and clothing. Edits are automatically propagated across all frames without regenerating the 3D geometry.

%% file: figures/fig_app.tex
\begin{figure}[hb]
    \centering
    \includegraphics[width=0.48\textwidth]{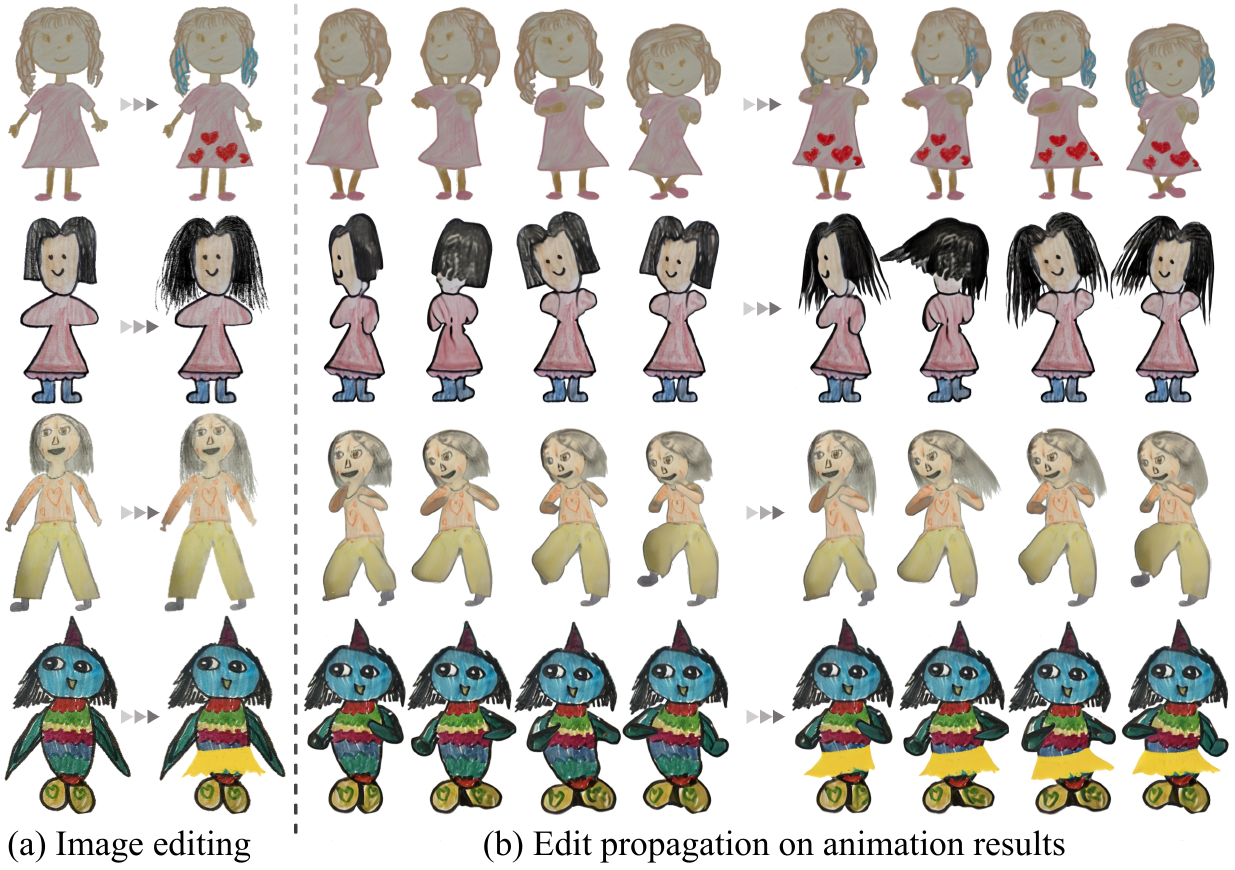}
    \vspace{-6mm}
    \caption{Edit propagation. The reference frame (left) is edited, and the change is propagated to all animation frames.}
    \vspace{-3mm}
    \label{fig:app}
\end{figure}

%% file: files/conclusion.tex
\section{Conclusion}
We introduced a hybrid system for animating hand-drawn characters with natural 3D motion by integrating skeletal animation and video diffusion priors. 
Our approach begins by generating coarse images via skeletal animation that maintain geometric consistency, followed by refinement using a domain-adapted diffusion model. \lz{Through our investigation of the denoising process in video diffusion models, we observed for the first time that different denoising steps are closely associated with distinct types of motion. This insight inspired the development of our novel Secondary Dynamics Injection (SDI) strategy.
The SDI strategy enhances motion realism in the refined images by guiding the denoising process with a pre-trained diffusion model, which leverages rich and realistic human motion priors.}
We also addressed unnatural deformation artifacts caused by the integrated hair-body single-mesh geometry with a Hair Layering Modeling (HLM) technique, enabling more natural animations for characters with long hair.
Our experiments demonstrate that the proposed method produces visually compelling 3D animations while preserving the artistic style and conveying natural secondary dynamics. 

\zj{In comparing our method to simulation-based secondary dynamics, it's important to note that the 3D models generated by our approach are rough geometries used as proxies, which do not meet the input requirements for simulation-based animation. Additionally, simulation methods are often labor-intensive and require specialized expertise, making them less accessible for novices. In contrast, our method prioritizes user-friendliness, enabling users with limited experience to create animations efficiently while tackling key challenges. The user-friendly editing applications also highlight the practicality and significance of our system.
}

\input{figures/fig_limitation}
\zj{
There are still some limitations in our method. First, while the system supports inverted poses through skeletal retargeting {(Fig. \ref{fig:limit} (a))}, the pre-trained human video diffusion model lacks sufficient priors for such motions, as they are rare in human video datasets. This limitation can lead to unrealistic secondary dynamics, such as hair failing to fall naturally under gravity.
Second, to prevent the model from learning unnatural deformations caused by the integrated hair-body geometry (Fig. \ref{fig:existing_problem}), such long-hair cases were excluded from training. However, this exclusion further limits the model's ability to generalize when inpainting hair regions in back views of long-hair characters. See the incorrect hair texture on the left shoulder in Fig. \ref{fig:limit} (b).
These two issues could potentially be addressed in the future by leveraging more powerful and realistic video generation models, capable of better handling complex motions.
Lastly, our current framework does not support explicit facial expression control, which deserves future exploration.
}

%% file: figures/fig_limitation.tex
\begin{figure}[h]
    \centering
    \includegraphics[width=0.48\textwidth]{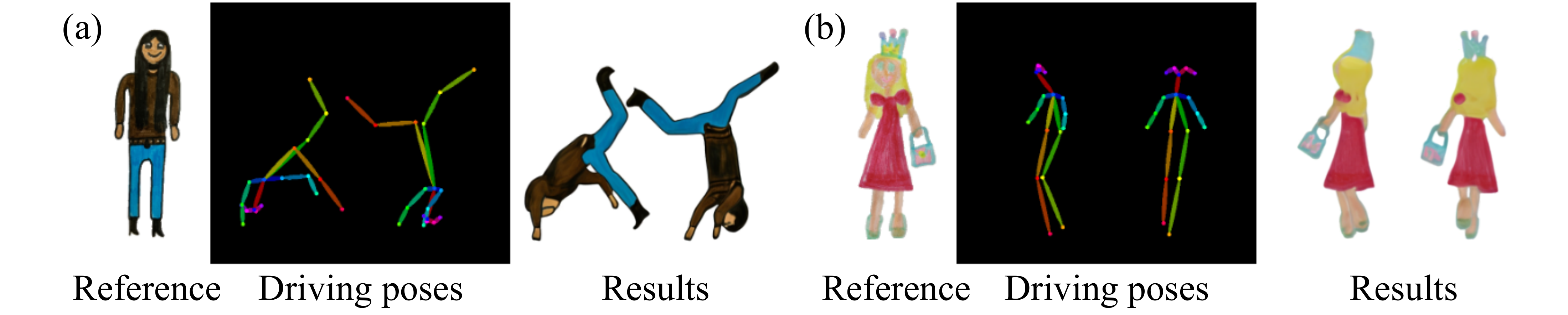}
    \vspace{-6mm}
    \caption{Two examples for the limitations of our method.}
    \vspace{-3mm}
    \label{fig:limit}
\end{figure}

%% file: figures/fig_SDI_mask_show.tex
\begin{figure*}[h]
    \centering
    \includegraphics[width=1\textwidth]{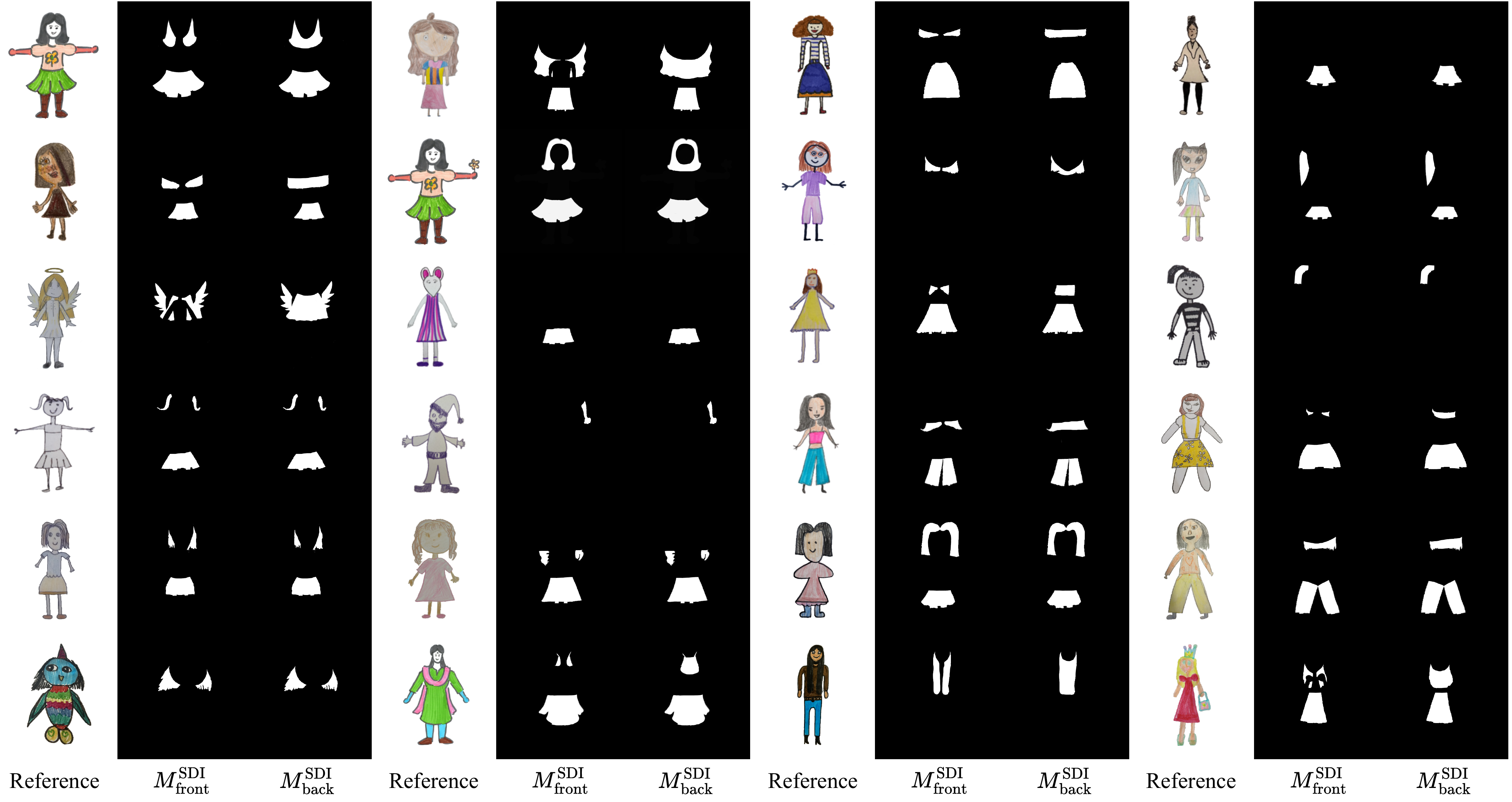}
    \vspace{-6mm}
    \caption{The SDI masks corresponding to all the examples used in the paper.}
    \label{fig:SDI_mask}
\end{figure*}